\documentclass[journal]{IEEEtran}
\ifCLASSINFOpdf
\else
\fi
\ifCLASSOPTIONcompsoc
 \usepackage[caption=false,font=normalsize,labelfont=sf,textfont=sf]{subfig}
\else
 \usepackage[caption=false,font=footnotesize]{subfig}
\fi
\usepackage[disable]{todonotes} 
\usepackage{hyperref}
\usepackage{amsmath}  
\usepackage{amssymb,mathtools,amsthm}
\usepackage{booktabs}
\usepackage{soul}
\usepackage{booktabs}
\usepackage{multirow}
\usepackage{siunitx}
\usepackage{colortbl}
\usepackage[inline]{enumitem}
\usepackage{threeparttable}  
\usepackage[export]{adjustbox}  
\usepackage{pgfplots}
\usepackage{tikz}  

\pgfplotsset{compat=1.17}

\setuptodonotes{inline}
\definecolor{babyblueeyes}{rgb}{0.63, 0.79, 0.95}
\definecolor{bittersweet}{rgb}{1.0, 0.44, 0.37}
\definecolor{caribbeangreen}{rgb}{0.0, 0.8, 0.6}
\definecolor{babypink}{rgb}{0.96, 0.76, 0.76}
\newcommand{\ljtodo}[1]{\todo[inline, color=bittersweet]{\normalfont@LJ #1}}

\newcommand{\jntodo}[1]{\todo[inline, color=caribbeangreen]{\normalfont@JN #1}}
\newcommand{\jncomment}[1]{\todo[inline, author=Jakub, color=caribbeangreen]{\normalfont#1}}
\newcommand{\thtodo}[1]{\todo[inline, color=babypink]{\normalfont@TH #1}}

\newcommand{\mstodo}[1]{\todo[inline, color=babyblueeyes]{\normalfont@MS #1}}

\newcommand{\E}[1]{\mathrm{E}[#1]}
\newcommand{\STD}[1]{\mathrm{STD}[#1]}

\hypersetup{
    colorlinks,
    linkcolor={red!80!black},
    citecolor={blue!80!black},
    urlcolor={blue!80!black}
}

\begin{document}
%
\title{Experiment Precision}
\title{Systematic Analysis of Experiment Precision Measures and Methods for Experiments Comparison}
%
%
%

\author{Jakub~Nawała,
        Tobias~Hoßfeld,
        Lucjan~Janowski,
        and~Michael~Seufert
\thanks{This work is licensed under the \href{http://creativecommons.org/licenses/by/4.0/}{Creative Commons Attribution 4.0 International License}. The research leading to these results has received funding from the
	Norwegian Financial Mechanism 2014-2021 under project 2019/34/H/ST6/00599.}%
\thanks{J. Nawała and L. Janowski are with AGH University of Science and Technology,
Institute of Telecommunications, Kraków, Poland
e-mail: jnawala@agh.edu.pl.}
\thanks{T. Hoßfeld and M. Seufert are with the University of Würzburg,
Institute of Computer Science, Würzburg, Germany.}}

%
%

\markboth{Submitted to IEEE Transactions on Multimedia}%
{Nawała \MakeLowercase{\textit{et al.}}: Systematic Analysis of Experiment Precision Measures and Methods for Experiments Comparison}
%



\maketitle

\begin{abstract}
The notion of experiment precision quantifies the variance of user ratings in a subjective experiment. Although there exist measures that assess subjective experiment precision, there are no systematic analyses of these measures available in the literature. To the best of our knowledge, there is also no systematic framework in the Multimedia Quality Assessment field for comparing subjective experiments in terms of their precision. Therefore, the main idea of this paper is to propose a framework for comparing subjective experiments in the field of MQA based on appropriate experiment precision measures. We present three experiment precision measures and three related experiment precision comparison methods. We systematically analyse the performance of the measures and methods proposed. We do so both through a simulation study (varying user rating variance and bias) and by using data from four real-world Quality of Experience (QoE) subjective experiments. In the simulation study we focus on crowdsourcing QoE experiments, since they are known to generate ratings with higher variance and bias, when compared to traditional subjective experiment methodologies. We conclude that our proposed measures and related comparison methods properly capture experiment precision (both when tested on simulated and real-world data). One of the measures also proves capable of dealing with even significantly biased responses. We believe our experiment precision assessment framework will help compare different subjective experiment methodologies. For example, it may help decide which methodology results in more precise user ratings. This may potentially inform future standardisation activities.
\jncomment{We should consider mentioning the link to our GitHub repo here.}
\end{abstract}


%
\IEEEpeerreviewmaketitle

\section{Introduction}
\label{sec:introduction}
%
%
%
%
\jntodo{Refer readers to the Margaret's framework, that also in some sense considers the precision of subjective experiments.}
\jncomment{We should consider mentioning the link to our GitHub repo in this section.}
\IEEEPARstart{I}{n} statistics, the term \emph{precision} quantifies how close measurements are to each other. This is often expressed as the reciprocal of the measurement variance. In this paper, we consider subjective experiments in the field of Multimedia Quality Assessment (MQA). Therefore, the measurements are subjective user ratings. These are simply opinions (expressed using a dedicated scale) of subjective experiment participants about stimuli presented to them (e.g., videos or images). In particular, the quantification of the user-perceived Quality of Experience (QoE) is based on such subjective experiments.

In this context, the term \emph{experiment precision} provides a measure that quantifies the variance of the user ratings across different stimuli in a subjective experiment. Typically, experiment precision measures are normalised in the range between 0 (highest possible precision, i.e., minimum variance) and 1 (minimum possible precision, i.e., maximum possible variance). We put forward three \emph{experiment precision measures}: 
\begin{enumerate*}
\item $g$ --- a Generalized Score Distribution (GSD) \cite{Janowski2019arxiv} based measure.
\item $\ell$ --- a measure based on the subject inconsistency parameter $\upsilon$ of the model presented in \cite{Li2020simple}. (We later refer to this model as the Li2020 model.)
\item $a$ --- based on the so-called ``SOS hypothesis'' \cite{hosseld2011sos}, where SOS stands for Standard Deviation of Opinion Scores.
\end{enumerate*}

\ljtodo{Please read the following paragraph and check whether you now agree with its contents.}
The main idea of this paper is to propose a framework for comparing subjective experiments in the field of MQA based on appropriate experiment precision measures. To this end, we present three \emph{experiment precision comparison methods}.
Each method checks whether there is a statistically significant difference between experiment precision measures coming from a pair of experiments. The goal of each method is to answer the following question: Is there a significant difference in experiment precision between a pair of experiments? (Please note that each experiment precision measure can be treated as a point estimator of experiment precision for a \textit{single} experiment. Each experiment precision method, on the other hand, checks for significant differences between a \textit{pair} of point estimators.)
To the best of our knowledge, we are the first in the MQA community to propose a systematic framework for comparing subjective experiments in terms of their precision. Although there are works proposing precision-related measures (e.g., \cite{hosseld2011sos,Pinson2020confidence}), they do not address the issue of formally comparing a pair of experiments.

We systematically analyse the three experiment precision measures and related experiment precision comparison methods. We do so both through a simulation study and by using real subjective user responses as well.
If it comes to the simulation study, we focus on subjective experiments conducted in a challenging crowdsourcing setting. This setting may lead to biases and uncertainty in the resulting ratings. On the one hand, \cite{varela2013increasing, hossfeld2014best,egger2017crowdsourcing,gardlo2012microworkers} observed a shift in user ratings toward higher scores (i.e., a positive user rating bias) in subjective experiments performed in a crowdsourcing environment. On the other hand, since crowdsourcing experiments often happen remotely, the general experiment environment may include hidden influence factors or other uncontrolled influence factors that negatively bias the user responses \cite{naderi2018speech,hossfeld2013best,sackl2016size,rao2021towards}. Furthermore, the crowdsourcing environment may cause high user rating uncertainty \cite{hossfeld2013best,hossfeld2014crowdsourcing}, which directly affects experiment precision (due to the increased variance of user ratings).
This is why, in our simulation study, we consider both positively and negatively biased user responses. We also take into account a large range of response uncertainty that reflects response variance observed in real subjective experiments.
Finally, although we specifically focus on crowdsourcing subjective experiments, we emphasise that our experiment precision measures and the related experiment precision comparison methods are applicable to any kind of subjective user study, i.e., conducted in a laboratory setting, via crowdsourcing, or through field trials.


\subsection{Research Questions} 
\label{ssec:research_questions}
In this paper, we answer the following research questions.
\begin{enumerate}[label=R\arabic*]
    \item \emph{Experiment precision measures}: Are the $g$, $a$, and $\ell$ experiment precision measures properly reflecting experiment precision? Are the measures able to capture user rating uncertainty and bias in a challenging crowdsourcing setting?
    \item \emph{Experiment comparison methods}: What are proper methods to compare the precision of two subjective experiments?
    \item \emph{Sensitivity of experiment precision comparison methods}: How are the experiment comparison methods behaving in different scenarios, taking into account a crowdsourcing setting with user uncertainty and user bias?
	\item \emph{Real-world application:} To what extent do various experiment precision measures produce the same verdict (e.g., both experiments are labelled as equally precise) for real-world subjective experiments? Are measure indications in line with expert intuition and domain knowledge regarding these real-world experiments?
\end{enumerate}

To answer research questions R1 to R3, the true user rating distribution for a single stimulus (e.g., a video) of a subject participating in a subjective experiment is required. 
In a subjective experiment, participants (a.k.a. subjects) provide a single numerical score (usually between 1 and 5; cf. clause 7.1.1 of ITU-T Rec. P.913~\cite{P913_2021}). There exist theories linking multiple such scores (a.k.a. responses) to an underlying user rating distribution (a.k.a. response distribution). For example, the work of Janowski and Pinson \cite{Janowski2015} introduces the concept of user bias (a.k.a. subject bias). They show that it is possible to extract from the subjective responses which subjects (and by how much) are systematically shifted in their responses, when compared to the average opinion of all other subjects. Building on top of this and similar works, we decide to perform a simulation study that assumes a certain user rating distribution generating model. In particular, we use a model taking into account subject bias and user rating diversity (which we also refer to as user rating uncertainty or response uncertainty). Thanks to this simulation study, we are able to systematically analyse how various experiment precision measures (and related experiment comparison methods) perform when we change subject bias and response uncertainty.

To answer research question R4, we consider selected experiments from four real-world subjective studies: one on Virtual Reality (VR) QoE, one on speech QoE, one on image QoE, and one on video QoE. Thanks to our domain knowledge and the existing literature on the topic, among other things, we know what the ordering should be of these four experiment types in terms of experiment precision. We compare this with the outcomes generated by our experiment precision measures.

We believe that our work is important to the field of MQA, as it proposes, to the best of our knowledge, the first framework that allows to formally compare subjective experiments in terms of experiment precision. An important application of such a framework is differentiating between various experiment methodologies. Our framework may help decide which methodology results in higher experiment precision. Such information may guide subjective methodology standardisation and help practitioners choose a methodology if high subjective responses precision is their top priority.


\subsection{Claims and Contributions}
\label{ssec:claims_contributions}
In this article, we make the following claims: 
\begin{enumerate}[label=(\roman*)]
    \item Our experiment precision measures allow to position a subjective experiment in relation to other experiment types (e.g., speech or video QoE experiments). They also allow to compare one experiment run with other runs having similar or modified setup. All of this done to compare experiment runs in terms of experiment precision. 
    \item The experiment comparison methods we introduce, provide a statistical test indicating whether experiment precision measures for a pair of experiments are statistically significantly different.
    \item Use the $\ell$ measure as a go-to measure to report experiment precision of your experiment.
\end{enumerate}

The following are our contributions that substantiate the claims we make.
\begin{enumerate}[label=C\arabic*]
    \item We put forward a notion of experiment precision.
    \item We introduce three experiment precision measures ($g$, $a$, and $\ell$) allowing to assess experiment precision of a single subjective experiment.
    \item We analyse experiment precision measures behaviour in a systematic simulation study reflecting a subjective experiment with user bias and user rating diversity.
	\item We suggest three novel methods to compare experiment precision between a pair of subjective experiments. Through a simulation study, we analyse methods behaviour and sensitivity to user bias and user rating uncertainty. 
    \item We test the three experiment precision comparison methods on real-world subjective data from experiments on VR, speech, image, and video QoE.
    \item We give guidelines regarding reporting experiment precision and argue why reporting experiment precision is beneficial for the research community. 
    \item We show that the measure $\ell$ (and the related experiment precision comparison method) works best both in our simulation study and when applied to real data.
    \item We make publicly available a GitHub repository (cf. \url{https://github.com/Qub3k/qoe-experiment-precision}). It contains the code allowing to easily run our experiment precision measures and the related experiment comparison methods. It also gives a chance to at least partially reproduce our results.
\end{enumerate}


\subsection{Structure of the Article} 
Section~\ref{sec:experiment_precision_measures} gives a detailed description of the notion of experiment precision. It also introduces the three experiment precision measures. Section~\ref{sec:methods} sets forth the three novel experiment precision comparison methods. Section~\ref{sec:simulation_study_methodology} describes the methodology behind our simulation study. Section~\ref{sec:results} provides the numerical results of our systematic analysis of the experiment precision measures and the related comparison methods. Section~\ref{sec:practical_use_case} shows how our experiment precision measures perform on real-world subjective data. Section~\ref{sec:discussion} discusses the results. Finally, Section~\ref{sec:conclusions_further_work} concludes the article and gives an outlook on future work.

\section{Notion of Experiment Precision and Measures of Its Assessment}
\label{sec:experiment_precision_measures}
Here, we first describe the notion of experiment precision in detail. Then, we present three experiment precision measures. Each of them stems from a model already described in the literature. Hence, we refer the reader to relevant research items. We also show how we use the existing models to arrive at each experiment precision measure.

\subsection{Notion of Experiment Precision}
\label{ssec:notion_of_experiment_precision}
When we speak of experiment precision, we mean the variance of the response generation process. For example, in the case of the synthetic data that we use in our simulation study, the precision of the experiment corresponds to the variance that we use in the data generator. There is an inverse relationship between experiment precision and data generator variance. The higher the data generator variance, the lower the experiment precision. (For more information on the data generator we use, please refer to Sec.~\ref{sec:simulation_study_methodology}.)

Naturally, the notion of experiment precision can be extended to real-life subjective responses as well. This time, contrary to the simulation study where experiment precision can be equated to the variance of the data generator, the precision of the experiment cannot be directly measured. Instead, a theorized response generating model has to be fitted to the observed responses. Using the parameters of the fitted model, one can then infer the experiment precision. Importantly, the models that are relevant in this context are models that separate bias (i.e., a constant shift in responses, see \cite{Janowski2014} and \cite{Janowski2015} for more details) from the variance. Although there are models that partition the variance into per subject, per stimulus or per distortion condition components \cite{Li2017, Li2020simple}, it is the total variance that is of our interest. This total variance corresponds to experiment precision. The previous statement also means that we do not treat changes in subject bias as changes in experiment precision. Differently put, having two experiments with the same total variance, but different biases, we treat them as having the same experiment precision. We point out, however, that this is a theoretical assumption, which does not hold in certain corner cases. For one thing, if subjective responses are provided on a discrete scale (which is often the case), then the change in mean response changes the variance. This is the case since all discrete domain probability distributions (and subjective responses can be treated as such \cite{Seufert2019d, Seufert2021}) have their mean and variance mutually dependent.

Experiment precision can be used for various reasons. The most obvious one is to report a new database of subjective responses. With experiment precision provided alongside the raw data, a prospective user of the database can quickly learn in what relation to other subjective experiment types this experiment is. For example, one can easily answer the following question: Are these data more or less precise than data coming from a typical video QoE experiment?
Furthermore, if an experiment is run in multiple sessions or locations, the notion of experiment precision can help make sure that all experiment runs are similar. Along with other indications, the notion of experiment precision could be used to decide whether the responses gathered in two experiment runs can be merged. Thanks to the information the notion of experiment precision provides, it could help decide which experiment setup results in more precise measurements as well.

\subsection{GSD Based Measure \textit{g}}
\label{ssec:gsd_based_measure}
\jntodo{$n \rightarrow N$ (i.e., revert back to Tobias' notation).}
\jntodo{$k \rightarrow K$ (i.e., revert back to Tobias' notation).}
Generalized Score Distribution (GSD) is a model describing subjective responses
generation process. Specifically, assuming subjective responses are treated
as realisations of a discrete random variable, the GSD represents a
family of discrete probability distributions. This family reflects distributions
observable in typical MQA experiments. Differently put, if a response
distribution can be described by the GSD model, there is a high chance that we are
dealing with properly conducted subjective experiment.

The GSD model was first introduced in~\cite{Janowski2019arxiv}. Model's
application to MQA subjective data and benefits related to using it were
later explored in~\cite{Nawala2020ACM} and~\cite{Nawala2022}. For the
concise description of the model, we refer the reader to~\cite{Nawala2022}.

The GSD represents per stimulus response distribution. The distribution is
parameterised with two parameters: $\psi$ and $\rho$. The first one ($\psi$) defines
the central tendency of the data and can be intuitively understood as
a drop-in replacement of the MOS measure.\footnote{Another intuitive description of the $\psi$ parameter is that it represents the mean opinion of the complete population of observers. In other words, it represents the MOS, as would be observed, if we asked about the opinion, all people, whose opinion we are interested in.} The second one ($\rho$) defines the spread of responses. It acts as a confidence parameter. Thus, the higher the $\rho$, the higher the confidence of people's opinions and, therefore, the lower opinions variability.

Since $\rho$ expresses opinions confidence, it is natural to associate it with
experiment precision. As the GSD is fitted per stimulus, there are as many
estimated values of $\rho$, as there are stimuli in a subjective experiment. In other words, if there are $K$ stimuli tested in the course of a subjective experiment, the GSD model is fitted $K$ times and we end up with $K$ estimates of $\rho$ (each denoted as $\hat{\rho}$). Now, to compute the $g$ measure we simply find the mean of the estimated $\rho$s, that is,
\begin{equation}
    g = \frac{1}{K} \sum_{i = 1}^{K} \hat{\rho}_i,
\end{equation}
where $K$ is the number of stimuli tested during the subjective experiment analysed and $\hat{\rho}_i$ is the estimated value of GSD's $\rho$ for the $i$-th stimulus.


\subsection{Li2020 Based Measure $\ell$}
\label{ssec:li2020_based_measure}
In~\cite{Li2020simple} Li \textit{et al.} introduce another model that represents subjective responses generation process. To make the discussion more comprehensible, let us refer to the model introduced in~\cite{Li2020simple} as the Li2020 model. The Li2020 model has three parameters. The three parameters correspond to: (i) true quality $\psi$ (conceptually similar to GSD's $\psi$), (ii) subject bias $\Delta$ (representing a systematic shift in the responses of a single subject, relative to the opinion of all the other subjects) and (iii) subject inconsistency $\upsilon$ (representing a random error). Although on the surface, the GSD and Li2020 models look similar, their internal structures differ significantly. For one thing, the Li2020 model uses an underlying continuous normal distribution (that is mapped to a discrete domain to reflect actually observed responses), whereas the GSD does not.\footnote{Readers interested in learning more about the differences between the GSD and Li2020 models are encouraged to take a look at~\cite{Nawala2022}.}

As mentioned, one of Li2020 model's parameters relates to subject inconsistency (the $\upsilon$ parameter). Intuitively, subject inconsistency must be related to experiment precision. Thus, we use this parameter to assess experiment precision.\footnote{Readers interested in learning about our motivation for choosing the subject inconsistency parameter $\upsilon$ as a basis of the precision measure $\ell$ are referred to Appendix~\ref{app:justification} (cf. the Supplemental material).} Since subject inconsistency is estimated on the per subject basis, we get as many estimated subject inconsistencies $\hat{\upsilon}$, as there are subjects taking part in a subjective experiment. Now, to arrive at the measure $\ell$, we compute the average estimated $\upsilon$. Assuming that $N$ subjects take part in the experiment, we can find the value of the measure $\ell$ as follows.
\begin{equation}
    \ell = \frac{1}{N} \sum_{i = 1}^{N} \hat{\upsilon}_i,
\end{equation}
where $\hat{\upsilon}_i$ is the estimated subject inconsistency $\upsilon$ for the $i$-th subject.

\jntodo{Explain why we chose $\epsilon$ as a parameter of the Li2020 that may be useful for measuring the precision of a subjective experiment.}


\subsection{SOS Hypothesis Based Measure \textit{a}}
\label{ssec:sos_hyp_based_measure}
The SOS hypothesis based experiment precision measure $a$ uses the SOS parameter $a$ of the experiment of interest. In turn, the SOS parameter $a$ is based on the so-called ``SOS hypothesis'' \cite{hosseld2011sos}. The SOS hypothesis states that in a typical QoE experiment, there is a simple quadratic relationship between mean opinion scores (MOS) and the standard deviation of opinion scores (SOS). 
The SOS values indicate per stimulus user rating diversity. A single parameter $a$ then relates the MOS and SOS values according to the quadratic relationship mentioned. The SOS parameter is bounded between 0 (no user rating diversity) and 1 (maximum user rating diversity). Hence, the SOS parameter directly reflects experiment precision.

Formally, we consider an experiment with $K$ test stimuli. 
Each stimulus is rated by all $N$ participants of that experiment on an Absolute Category Rating (ACR) scale \cite{itut_p910}, ranging from bad (1) to excellent (5) quality.
This allows to obtain the MOS $m_x$, i.e., the mean of all ratings, and the rating variance $v_x$ of each stimulus $x=1,\dots,K$.
The SOS hypothesis states that in typical QoE experiments, there is a simple quadratic relationship between the MOS $m$ and rating variance $v$ of all stimuli. Hence, the rating variance is a quadratic function of $m$ with a constant value $a$, referred to as the SOS parameter. The function has the following form for a 5-point rating scale:
\begin{align}
 &v = f_a(m) = a\cdot(5-m)\cdot(m-1) \;, & 1 \leq m \leq 5 \;. \label{eq:mossos}
\end{align}

For calculating the SOS parameter $a$ of an experiment, we take the MOS values $m_x$ and rating variances $v_x$ of all stimuli of that experiment and fit the corresponding MOS--SOS curve to obtain parameter $a$. 
The SOS parameter can be directly computed, see \cite{hossfeld2016formal}, via ordinary least-squares (OLS) regression:
\begin{align}
 & a = \frac{\sum_{x=1}^{K} (5-m_x) \cdot (m_x-1) \cdot v_x}{\sum_{x=1}^{K} (5-m_x)^2 \cdot (m_x-1)^2}. \label{eq:sos}
\end{align}
We directly use the $a$ parameter computed this way as an indication of experiment precision (referring to the approach as the experiment precision measure $a$).




\section{Experiment Precision Comparison Methods}
\label{sec:methods}
\jntodo{Consider merging this section with Sec.~\ref{ssec:notion_of_experiment_precision}.}
In this section, we describe three experiment precision comparison methods. They allow to assess whether there is a statistically significant difference in terms of experiment precision between a pair of subjective experiments. Importantly, the methods are based on the three experiment precision measures.

\subsection{Comparison Method Based on the \textit{a} Measure}
\label{ssec:pair_comp_method_a_measure}
We start by computing the SOS parameters $a_1$ and $a_2$ for the two subjective experiments being compared. For this, we use the procedure described in Sec.~\ref{ssec:sos_hyp_based_measure}. Having two parameter estimates, we compare them using the independent two-sample $t$-test (assuming unequal sample variances and unequal sample sizes). The null hypothesis is that the two $a$ parameters are the same. Since we use the 5\% significance level, the null hypothesis is rejected only if the resultant $p$-value is less than or equal to 0.05. Please note that since we assume that the estimated SOS parameter $a$ is equivalent in value to the experiment precision measure $a$, the procedure described effectively compares $a$ measures between a pair of experiments. We refer to Appendix~\ref{app:comp_of_two_a_measures} (cf. the Supplemental material) readers interested in details regarding how two SOS parameters $a$ can be formally compared with the $t$-test.


\subsection{Comparison Method Based on the \textit{g} Measure}
\label{ssec:pair_comp_method_g_measure}
As we already mentioned in Sec.~\ref{ssec:gsd_based_measure}, the GSD model is estimated on the per stimulus basis. Thus, there are as many
estimated values of $\rho$, as there are stimuli in a subjective experiment.
Let us denote such a vector of $\rho$ estimates as $\boldsymbol{\hat{\rho}}$.
Having two such vectors from a pair of experiments we wish to compare
($\boldsymbol{\hat{\rho}}_1$ and $\boldsymbol{\hat{\rho}}_2$), we apply
a two-sample independent $t$-test on them, assuming unequal variances in the two
samples. The null hypothesis is that the two vectors have the same average value. In other words, the null hypothesis states that the two experiments have the same value of the $g$ measure. Finally, we use \textit{t}-test's $p$-value as an indication of whether the two experiments differ significantly in terms of precision.


\subsection{Comparison Method Based on the $\ell$ Measure}
\label{ssec:pair_comp_method_l_measure}
As Li2020's subject inconsistency $\upsilon$ is estimated on the per subject basis, we get as many estimated subject inconsistencies $\hat{\upsilon}$, as there are subjects taking part in a subjective experiment. Let us denote this vector of estimated subject inconsistencies as $\boldsymbol{\hat{\upsilon}}$. Having two such vectors from a pair of experiments we wish to compare ($\boldsymbol{\hat{\upsilon}}_1$ and $\boldsymbol{\hat{\upsilon}}_2$), we apply a two-sample independent $t$-test on them, assuming unequal variances in the two samples. The null hypothesis is that the two vectors have the same average value. Differently put, the null hypothesis is that the two experiments have the same value of the $\ell$ measure. We use $t$-test's $p$-value to assess whether the two experiments differ significantly in terms of experiment precision.


\section{Simulation Study Methodology}\label{sec:simulation_study_methodology}

This section describes the simulation study. It also details how our data generator is structured and what parameter set we use to generate the data that we later utilize in Sec.~\ref{sec:results}.

\subsection{User Ratings as a Random Variable}
\label{ssec:user_ratings_as_a_rv}
In this work, we consider discrete subjective quality assessment rating scales. We particularly focus on the common 5-point ACR scale (5: Excellent, 4: Good; 3: Fair, 2: Poor, and 1: Bad; cf. clause 7.1.1 of ITU-T Rec. P.913~\cite{P913_2021}). For any stimulus (or test condition) and any user, we assume that user ratings (a.k.a. subjective responses) are realizations of a random variable $Q$. Notably, $Q \sim F(\lambda, \theta)$, where $F(\lambda, \theta)$ is a discrete distribution with mean $\lambda$ and standard deviation $\theta$. Importantly, $\lambda$ represents the central tendency of subjective responses, while $\theta$ quantifies their spread (which can be intuitively related to per subject response uncertainty).

The subject bias theory \cite{Janowski2015} proposes to derive $F(\lambda, \theta)$ from a normal distribution $\mathcal{N}(\mu, \sigma)$. Specifically, the authors of \cite{Janowski2015} censor the range of $\mathcal{N}(\mu, \sigma)$ to the interval $\left<1,5\right>$ and then discretize the distribution by rounding its values to the nearest integer. The censored and discretized continuous normal distribution leads to the so-called \emph{ordered probit} distribution \cite{Becker1992} with parameters $\mu$ and $\sigma$:
\begin{align}\label{eq:qnorm}
    Q \sim \mathcal{O}(\mu, \sigma) &= \text{round}\left( \left[ \mathcal{N}(\mu, \sigma) \right]_1^5 \right) \nonumber\\
    &= \mathrm{QNorm(\mu, \sigma)},
\end{align}
where $\mathcal{O}$ represents the ordered probit distribution, $\text{round}()$ represents discretisation and $\left[ \, \right]_1^5$ represents censoring. Importantly, the ordered probit distribution can also be thought of as a quantised version of the normal distribution. Thus, some authors refer to it as the \textit{quantised normal} distribution or simply $\mathrm{QNorm}$. We stick to this convention and refer to the ordered probit distribution as $\mathrm{QNorm}$ (or $\mathrm{QNormal}$).

In practise, to arrive at the $\mathrm{QNorm}$ distribution, one integrates the cumulative distribution function (CDF) $\Phi$ of a normal distribution $\mathcal{N}(\mu, \sigma)$ using a set of cut-off points (a.k.a. thresholds). Thus, the probability of observing each of the five ACR response categories is as follows:
\begin{align}
    p_1 & = \Phi(1.5), \nonumber \\
    p_i & = \Phi(i + 0.5) - \Phi(i - 0.5), \quad i = 2, 3, 4, \nonumber \\
    p_5 & = 1- \Phi(4.5),
\end{align}
where $p_i = P(Q = i)$.
The response category probabilities defined this way constitute the bulk of our synthetic data generator.

Please note that, in general, the expected user rating $\E{Q} = \sum_{i=1}^5 i \cdot p_i$ does \textit{not} equal $\mu$. In other words, if we were to artificially generate responses using (\ref{eq:qnorm}) and then compute their mean, this mean would not equal the value of $\mu$ we used to generate the responses in the first place. The same is true for the standard deviation, i.e., $\STD{Q} \neq \sigma$. These two effects occur due to discretisation and censoring mentioned before.

\begin{figure}[ht]
    \centering
    \includegraphics[width=\columnwidth]{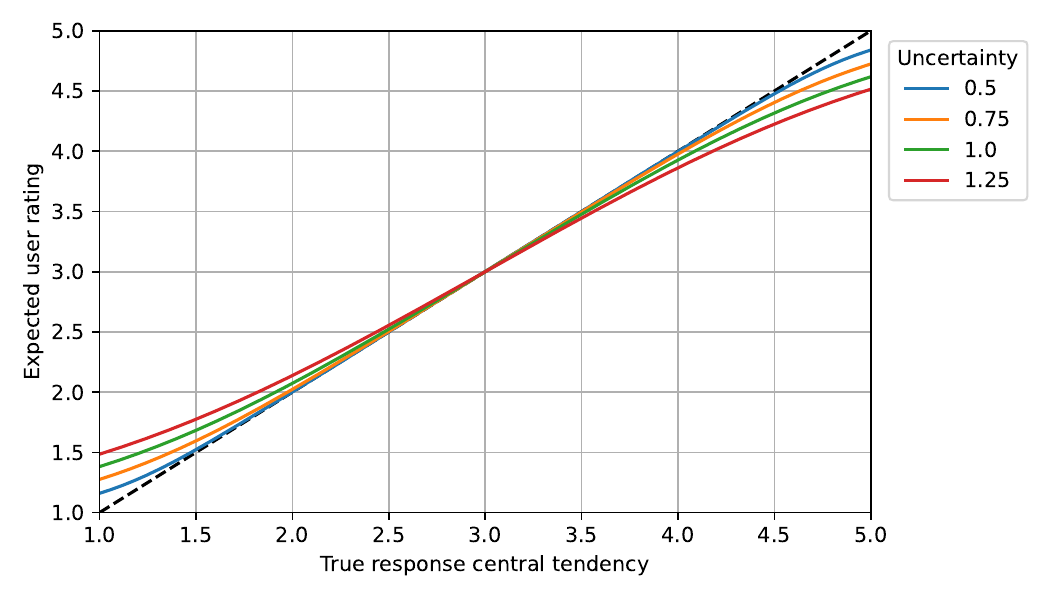}
    \caption{Expected user rating $\E{Q}$ vs data generator provided uncertainty $\sigma$ and true response central tendency $\mu$ for the $\mathrm{QNorm}$ model (cf. (\ref{eq:qnorm})).}
    \label{fig:inputMean2expected}
\end{figure}

\begin{figure}[ht]
    \centering
    \includegraphics[width=\columnwidth]{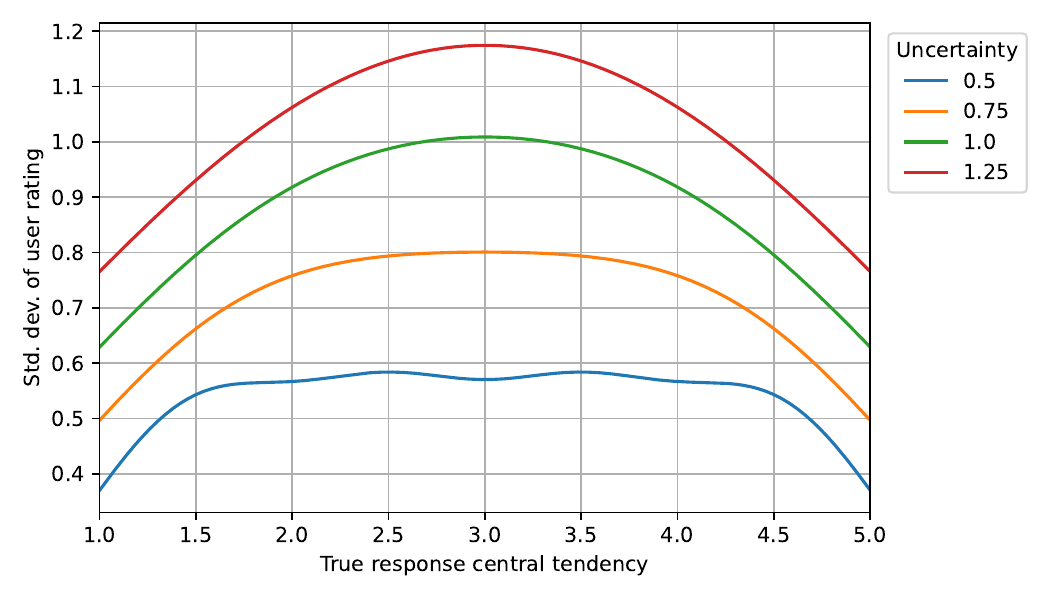}
    \caption{Standard deviation $\STD{Q}$ of the user rating distribution vs data generator provided uncertainty $\sigma$ and true response central tendency $\mu$ for the $\mathrm{QNorm}$ model (cf. (\ref{eq:qnorm})).}
    \label{fig:inputMean2std}
\end{figure}

Figure~\ref{fig:inputMean2expected} shows the mismatch between $\E{Q}$ and $\mu$ for multiple uncertainty values $\sigma$. Please note that the mismatch is greater at the limits of $\mu$'s value range than it is for $2\leq \mu \leq 4$. Figure~\ref{fig:inputMean2std} presents the mismatch between $\STD{Q}$ and $\sigma$.

Importantly, the standard deviation $\STD{Q}$ is derived from the following equation:
\begin{align}
\STD{Q} = \sqrt{\E{Q^2} - \E{Q}^2}
\end{align}
with $\E{Q^2} = \sum_{i=1}^5 i^2 \cdot p_i$.
Due to discretisation, the maximum standard deviation $\STD{Q}_{\max{}}$ may be greater than $\sigma$. This is the case for $\sigma \leq 1.0$. Please note that $\STD{Q}_{\max{}}$ usually occurs for $\mu=3$. For example, for $\sigma=0.75$, $\STD{Q}_{\max{}} \approx 0.8$. Please consider that at the limits of $\mu$'s value range, $\STD{Q}$ is smaller than $\sigma$. This happens because of the normal distribution censoring mentioned before.

We refer to \cite{Nawala2022} readers interested in learning more about the mismatch between $\mu, \sigma$ and $\E{Q}, \STD{Q}$, respectively. The authors of \cite{Nawala2022} detail how this mismatch influences subjective responses analysis and how to deal with this issue.


\subsection{Bias Due to Crowdsourcing}
Performing an MQA experiment in the crowdsourcing setting can lead to biases in the resulting ratings. To take this into account, we assume a varying prevalence of bias (cf. the discussion about the $p$ parameter in Sec.~\ref{ssec:simulation_parameters_and_scenarios}).

If a user is not biased, we express their ratings as
\begin{align}
Q \sim \mathrm{QNorm}(\mu, \sigma).
\end{align} 
However, for a biased user, we shift the central tendency of their ratings by $\beta$.
\begin{align}
Q \sim \mathrm{QNorm}(\mu + \beta, \sigma)
\end{align}

Based on our domain knowledge and the existing literature on the topic, we differentiate between a positive and negative bias.
 
\paragraph{Positive bias} In crowdsourcing studies, we observe a shift of user ratings toward higher ratings for different reasons. One of them is the willingness to pleasure the employer \cite{varela2013increasing, hossfeld2014best,egger2017crowdsourcing}. A strong bias of paid users to rate images towards the top-end of the quality scale was also observed in \cite{hossfeld2014best,gardlo2012microworkers}. These and similar effects may be taken into account by shifting the mean $\mu$ of the underlying normal distribution by a constant $\beta>0$. Typical values for $\beta$ are in the order between 0.25 and 0.75 \cite{volk2015crowdsourcing,hossfeld2013best,hossfeld2014crowdsourcing,hossfeld2014best}.

\paragraph{Negative bias}
There may be users who cannot properly consume the test content, e.g. due to the noisy environment \cite{naderi2018speech}, hidden influence factors (e.g., improper device \cite{hossfeld2013best}), or different cultural perceptions of aesthetics \cite{varela2013increasing}. The authors of \cite{sackl2016size} explain that ``undesired side effects like fatigue or boredom could have caused the lower ratings, i.e., crowdsourcing test participants were more and more annoyed.'' Such a negative bias was found across different types of stimuli and experiments, e.g., in video QoE experiments \cite{rao2021towards}. We take such effects into account by shifting the mean $\mu$. However, this time we use a negative value of $\beta$. To our knowledge, the realistic range for $\beta$ is between $-0.75$ and $-0.25$.


\subsection{Simulation Framework}
\begin{table}
 \centering
 \begin{threeparttable}
 \caption{Naming Convention and Simulation Study Parameters} 
 \label{tab:notation_and_sim_config}
 \begin{tabular}{ll}
 \toprule
   Variable  & Description \\
 \midrule
     $k$ & number of stimuli in a subjective experiment [$k=21$]\\[0.5em]
     $n$ & number of users rating $k$ stimuli [$n=30$]\\[0.5em]
     $\mu_x$ & true average user rating for stimulus $x=1,\dots,k$\\
     & [$\mu_x \in \{1, 1.2, 1.4, 1.6, \dots, 4.4, 4.6, 4.8, 5\} $] \\[0.5em]
     $\beta_{u}$ & bias of user $u$ [$\beta_u \in \left\{ \pm1, \pm0.5 \right\}$] \\
                & (under extreme or mixed symmetric bias scenario, resp.)\\[0.5em]
     $\sigma_{u}$ & uncertainty of user $u$ \\   
     & [$\sigma \in \{0.4 , 0.45, 0.5 ,\dots, 1.15, 1.2 , 1.25\} $] \\[0.5em]
			$p$ & no bias probability \\
			& [$p \in \{0, .1, .2, .3, .4, .5, .6, .7, .8, .9, .98, 1\}$]\\[0.5em]
	$Q_{u,x}$ & discrete user rating distribution of user $u$ for stimulus $x$ \\
	& [$Q_{u, x} \sim \mathrm{QNorm}(\mu_x + \beta_{u}, \sigma_u) $]\\
 \bottomrule
 \end{tabular}
 \begin{tablenotes}
    \item Simulation study parameters are given in square brackets.
 \end{tablenotes}
 \end{threeparttable}
\end{table}

We are simulating individual QoE experiments. Each experiment consists of a fixed number $n$ of users, who are rating $k$ stimuli (or test conditions).
For each user $u=1,2,\dots,n$, the bias $\beta_u$ and the uncertainty $\sigma_u$ of that user $u$ are known.
For each stimulus, there is a true average user rating $\mu_x$ (without subject bias).

To model user responses coming from a single crowdsourcing QoE experiment, we make the following assumptions.
 \begin{enumerate}
    \item The bias $\beta_u$ of a user $u$ is constant throughout the experiment across different stimuli. 
    \item The uncertainty $\sigma_u$ of a user $u$ is constant throughout the experiment across different stimuli. Importantly, we assume that by changing $\sigma_u$ we are effectively changing the experiment precision of a simulated experiment.
    \item The simulated QoE experiment consists of $k$ stimuli (or test conditions) that are evaluated by $n$ subjects. 
	\item For each stimulus $x$, there is a true average user rating $\mu_x$ (without subject bias). We also assume a perfect selection of stimuli, which results in equidistant values of $\mu_x$, all in the range from 1 to 5. Differently put, $\mu_x = 1 + (x-1)\frac{5-1}{k-1}$ for $x=1,\dots,k$.
    \item Finally, for stimulus $x$ and user $u$, the ratings $Q_{u,x}$ follow the $\mathrm{QNorm}$ distribution with parameters $\mu_x+\beta_u$ and $\sigma_u$:
		\begin{align}
		Q_{u,x} \sim \mathrm{QNorm}(\mu_x+\beta_u, \sigma_u). \label{eq:data_generator}
		\end{align}
\end{enumerate}
Equation~(\ref{eq:data_generator}) completely describes our data generator. Using it, we randomly draw responses for user $u$, who assesses stimulus $x$. We also randomly select the related subject bias $\beta_u$. The next section details how this is done.


\subsection{Simulation Parameters and Scenarios}
\label{ssec:simulation_parameters_and_scenarios}
\begin{figure}[ht]
    \centering
    \includegraphics[width=\columnwidth,clip,viewport=0cm 0.5cm 17.78cm 9.8cm]{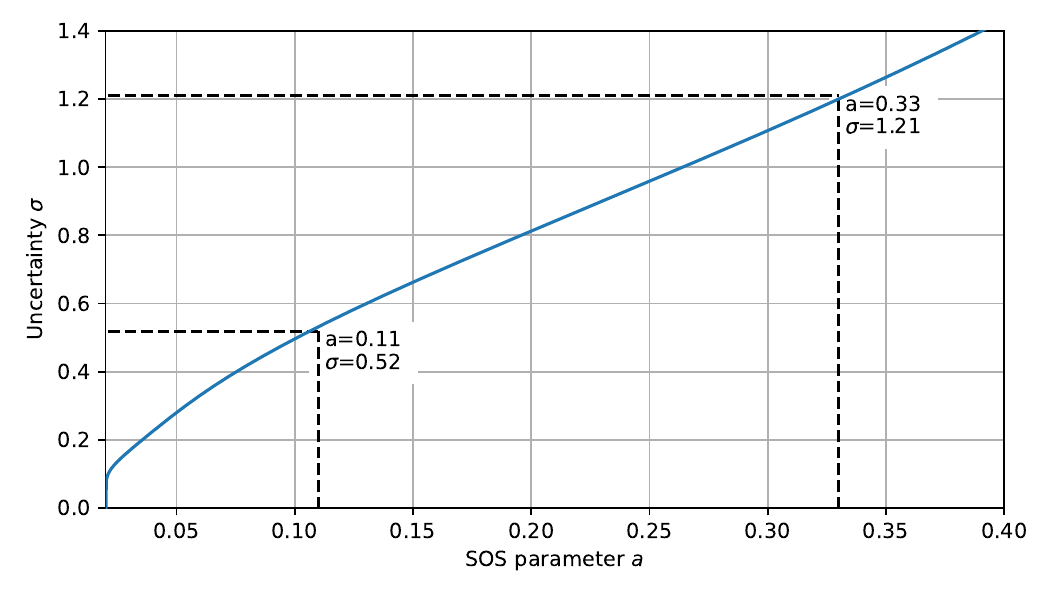}
    \caption{Selection of uncertainty values $\sigma$ based on SOS results \cite{hosseld2011sos}.
    }
    \label{fig:mappingSOSsigma}
\end{figure}

In our simulation study, we assume that the pool of subjects in the experiment has the same fixed uncertainty $\sigma=\sigma_u$ for any $u$. Hence, the uncertainty depends only on the application or service under investigation. Interestingly, the uncertainty $\sigma$ of the $\mathrm{QNorm}$ model corresponds to a dedicated SOS parameter $a$. In \cite{hosseld2011sos}, the SOS parameter $a$ was quantified for various applications (in the context of QoE subjective experiments). Each QoE experiment was assigned a single SOS parameter. The range of SOS parameters spanned between $a=0.11$ and $a=0.33$. Figure~\ref{fig:mappingSOSsigma} shows the mapping between $\sigma$ and SOS parameter $a$ for simulated experiments spanning the same range of SOS parameters as the one reported in \cite{hosseld2011sos}.\footnote{To create this plot, we considered $k=21$ stimuli with equidistant $\mu_x$ values, i.e., $\mu_x \in \{1 , 1.2, 1.4, \dots, 4.6, 4.8, 5\}$. We also assumed no bias, i.e., $\beta=0$. Hence, the user rating distribution $Q_{u,x} \sim \mathrm{QNorm}(\mu_x,\sigma)$ is the same for each user $u$. Having generated the simulated responses, we obtain the MOS value $\E{Q_{u,x}}=\hat{\mu}_x$ and the SOS value $\STD{Q_{u,x}}=\hat{\sigma}_x$, both for each stimulus $x$. Based on the MOS--SOS tuples ($\hat{\mu}_x,\hat{\sigma}_x$), we estimate the SOS parameter $a$ (cf.~\cite{hossfeld2016formal}).} Thanks to this mapping, we are able to use a range of $\sigma$ values that corresponds to real-life subjective QoE experiments.

We now describe in detail the parameter set we use to perform the simulation study.
If it comes to uncertainty $\sigma$, we vary it in the range $\left<0.4, 1.25\right>$ with a step size of $0.05$: 
$
\sigma \in \{0.4 , 0.45, 0.5 ,\dots, 1.15, 1.2 , 1.25\}.
$
We consider a fixed number of $k=21$ stimuli which have a true average user rating:
$
\mu_x \in \{1, 1.2, 1.4, 1.6, \dots, 4.4, 4.6, 4.8, 5\}.
$

For simulating subject bias, we use a mixed symmetric bias approach (more on that later). With a certain probability $p$ a user is \textit{not} biased.
Therefore, with probability $1-p$ users do have a bias. In the simulation study, we consider the following 12 values of ``no bias probability'':
$
p \in \{0, .1, .2, .3, .4, .5, .6, .7, .8, .9, .98, 1\}.
$
At the beginning of each simulated experiment, we draw a random number from a Bernoulli experiment with parameter $p$ to determine if a user is not biased. Please note that the bias $\beta_u$ is then fixed for that user for all stimuli $x=1,\dots,k$. Moreover, note that the number of users without bias follows a binomial distribution with parameters $n$ and $p$. The expected number of users without bias is $n\cdot p$. Significantly, we only use simulated experiments with $n=30$ subjects.


We consider the mixed symmetric bias approach in two scenarios: i) mixed symmetric bias and ii) extreme symmetric bias.
In the mixed symmetric bias scenario, we assume that with a probability $p$ a user has no bias. With probability $(1-p)/2$, a user has a positive bias $\beta_u=0.5$. Similarly, with probability $(1-p)/2$, a user has a negative bias $\beta_u=-0.5$. The following equation describes this approach formally.
\begin{align}
\beta_u = 
\begin{cases}
-0.5 & \text{with prob. } (1-p)/2, \\
	0 & \text{with prob. } p, \\
	+0.5 & \text{with prob. } (1-p)/2.
\end{cases}
\end{align}

In the extreme symmetric bias scenario, we assume that half of the users have a positive bias ($\beta_u=+1$) and the other half have a negative bias ($\beta_u=-1$). This scenario is intended to identify the limits of the proposed experiment precision measures. We call this scenario extreme, since all subjects have a strong bias, with a bias level atypical of regular subjective experiments. The following equation describes this scenario formally.
\begin{align}
\beta_u = 
\begin{cases}
-1 & \text{with prob. } 1/2, \\
	+1 & \text{with prob. } 1/2.
\end{cases}
\end{align}
Please note that the extreme symmetric bias scenario may serve as an example of a subjective experiment, where the pool of participants is divided in half. One half has a positively skewed opinion about the stimuli presented. The other half has their opinion negatively skewed. Although it is difficult to talk about the consensus regarding stimulus quality in such a group of participants, it is interesting to check how our experiment precision measures respond to such conditions.

Each experiment configuration (i.e., one selected bias scenario, no bias probability $p$, and global uncertainty $\sigma$) is repeated $r=200$ times to get statistically significant results. Table~\ref{tab:notation_and_sim_config} concisely presents the set of simulation parameters we use. It also demonstrates our naming convention. We refer to Appendix~\ref{app:data_gen_conf_influence} (cf. the Supplemental material) readers interested in learning more about how the parameters we provide to our data generator influence the responses generated.


\section{Results}
\label{sec:results}
In Sec.~\ref{ssec:systematic_analysis} we provide a systematic analysis of the three precision measures. The section presents how the measures respond to changes in bias probability ($p$) and response uncertainty ($\sigma$). Such analyses highlight the applicability scope of each measure. In Sec.~\ref{ssec:comparisons_of_experiments} we present how our experiment comparison methods (internally using the experiment precision measures) perform on the simulated data. Among others, we quantify how well each method differentiates between experiments that have different experiment precision.
    

\subsection{Systematic Analysis of Models Underlying Experiment Precision Measures}
\label{ssec:systematic_analysis}
We now show how the models underlying our experiment precision measures respond to changes in response bias probability ($1-p$) and response uncertainty ($\sigma$).

\begin{figure*}[!t]
    \centering
    \subfloat[]{\resizebox{!}{1.45in}{\includegraphics[clip, viewport=0 0.35cm 15cm 10cm]{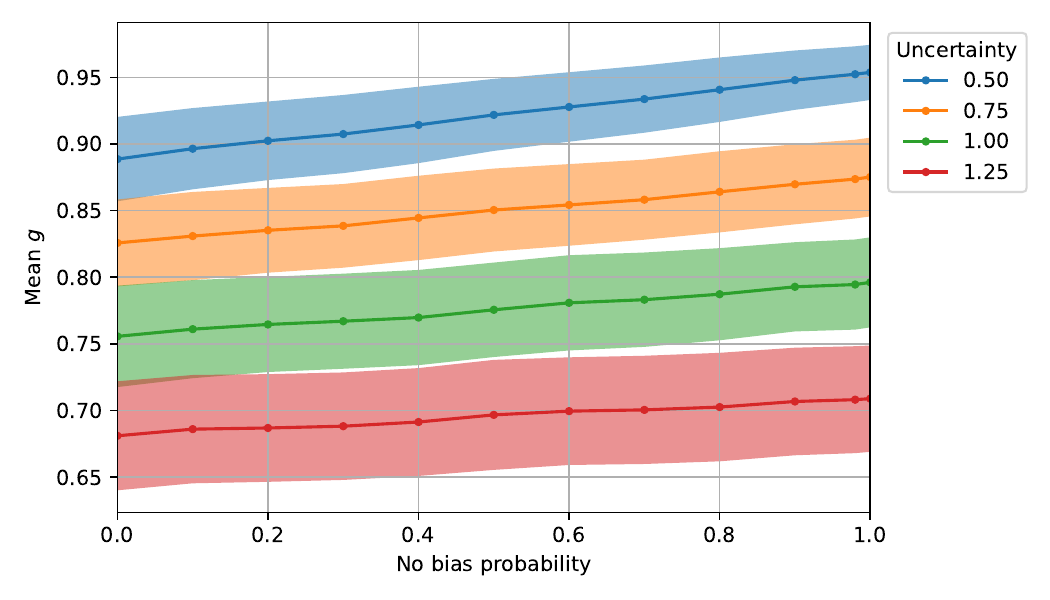}}%
        \label{fig:rho_bar_vs_no_bias_prob}}
    \hfil
    \subfloat[]{\resizebox{!}{1.45in}{\includegraphics[clip, viewport=0 0.35cm 15cm 10cm]{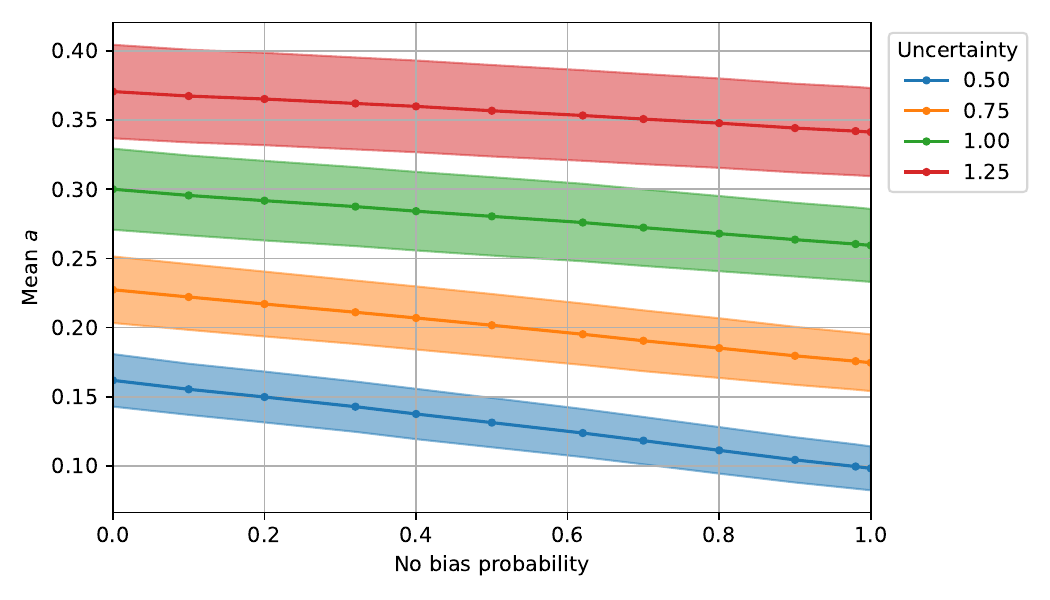}}%
        \label{fig:sosa_vs_no_bias_prob}}
    \hfil
    \subfloat[]{\resizebox{!}{1.45in}{\includegraphics[clip, viewport=0.5cm 0 17.1cm 9.15cm]{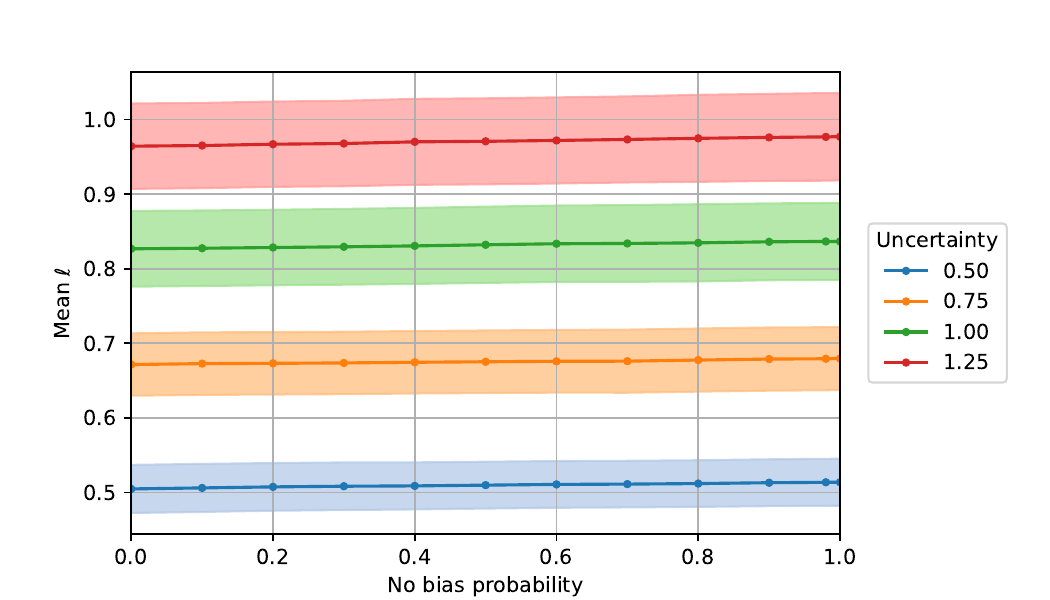}}%
        \label{fig:upsilon_bar_vs_no_bias_prob}}
    \caption{Results of fitting a model parameter to the simulated data for varying no bias probability $p$ and uncertainty $\sigma$. The shaded area corresponds to the average per experiment 95\% confidence interval. (Refer to Appendix~\ref{app:confidence_intervals} in the Supplemental material to learn how we compute the confidence interval for each measure.) Display (a) corresponds to the $g$ measure, display (b) to the $a$ measure, and display (c) to the $\ell$ measure. Please note that results in (b) and (c) use all 200 runs of the data generator, whereas (a) presents results for only 20 runs.}
    \label{fig:measure_vs_no_bias_prob}
\end{figure*}

\jncomment{It looks pretty suspicious that we decided to run the GSD model for only 20 runs of the data generator. I have a feeling we should repeat the analysis with all 200 runs. (Especially so, since we now have the hardware accelerated estimation of GSD parameters available.)}
Let us start by investigating the GSD model. To test the model, we fit it to responses from the first 20 runs of our data generator.\footnote{We perform the analysis for 20 runs only (out of 200 runs in total), since the estimation process for the GSD is computationally demanding.} Importantly, we only investigate the following uncertainties $\sigma$: $\left\{ .5, .75, 1, 1.25 \right\}$. Having fitted the model to each stimulus in the simulated experiments, we compute a per experiment mean estimated $\rho$ (and denote it as $g$). To learn how much $g$ can change across different data generator runs, we also show the 95\% confidence intervals of the computed $g$ values.\footnote{Refer to Appendix~\ref{app:confidence_intervals} in the Supplemental material to learn how we compute the confidence intervals.} Fig.~\ref{fig:rho_bar_vs_no_bias_prob} presents the results of this analysis.
As we can see, the $g$ measure behaves properly. What is more, the greater the uncertainty, the smaller the value of $g$. (Please keep in mind that $g$ is a measure of confidence and thus has an inverse relationship to variance.) Furthermore, the higher the bias probability (i.e., $1-p$, where $p$ represents no bias probability), the lower the value of $g$ (for a fixed value of uncertainty $\sigma$). This means that the response bias influences the value of $g$. Since we would like to ideally have an experiment precision measure not sensitive to changes in response bias, this feature of $g$ measure is undesirable. At last, please note that the estimation of $g$ is more precise (i.e., the confidence interval is narrower) for low uncertainty than it is for high uncertainty.

We now repeat the analysis for the SOS $a$ model. This time, we use all 200 runs of the data generator. Fig.~\ref{fig:sosa_vs_no_bias_prob} presents the results of this analysis. As in the case of the GSD model, the $a$ parameter responds to changes in uncertainty $\sigma$. The higher the uncertainty, the greater the value of $a$. This is a positive feature of $a$. (Keep in mind that the theory underlying the $a$ parameter states that high values of the $a$ parameter correspond to imprecise experiments.) However, Fig.~\ref{fig:sosa_vs_no_bias_prob} also reveals that $a$ is sensitive to changes in bias probability ($1-p$). This is \textit{not} desirable. As the bias probability increases (this corresponds to moving horizontally from the RHS to the LHS of the figure), the value of $a$ increases. This suggests that the $a$ parameter interprets the increasing response bias as a larger and larger uncertainty. In general, the behaviour of the SOS $a$ model is quite similar to the behaviour of the GSD model.

Fig.~\ref{fig:upsilon_bar_vs_no_bias_prob} presents the results of fitting the Li2020 model to the simulated experiments. The fitting is repeated 200 times for each data generator configuration (i.e., a fixed value of no bias probability and a fixed value of uncertainty). For each simulated experiment, we estimate subject inconsistency $\upsilon$ for each subject. Then, we take the average estimated subject inconsistency in the scope of a single experiment and denote it as $\ell$. Finally, we average the $\ell$ measure over 200 data generator runs (for a given data generator configuration). Fig.~\ref{fig:upsilon_bar_vs_no_bias_prob} also shows the mean confidence interval (computed over 200 runs per data generator configuration). In general, the $\ell$ measure seems to work properly. Changing the uncertainty results in changes in the measure $\ell$. The higher the uncertainty $\sigma$, the higher the value of $\ell$. This is a positive feature of the $\ell$ measure. Note that, similarly to the $g$ measure, the estimation of the $\ell$ measure is more precise for low uncertainty than for high uncertainty (cf. confidence interval width). Unlike the $g$ and $a$ measures, the $\ell$ measure is generally \textit{not} sensitive to changes in bias probability. This is also a positive feature of the $\ell$ measure. Furthermore, this feature has a special meaning, since it is directly in line with our definition of experiment precision. More generally, the $\ell$ measure is the only measure that closely follows our definition of experiment precision. Thus, we expect the $\ell$ measure to perform well as an experiment precision estimator.

Please take notice of the spacing between consecutive lines in Fig.~\ref{fig:rho_bar_vs_no_bias_prob} and Fig.~\ref{fig:upsilon_bar_vs_no_bias_prob}. Note that this spacing is constant in Fig.~\ref{fig:rho_bar_vs_no_bias_prob}. The same is \textit{not} true in Fig.~\ref{fig:upsilon_bar_vs_no_bias_prob}. In other words, for the $g$ measure, comparing pairs of experiments with consecutive uncertainty values (e.g., $0.50$ \& $0.75$ or $0.75$ \& $1.00$) has a more or less constant chance of detecting significant differences. However, in the case of the $\ell$ measure, it is more probable to detect significant differences when comparing experiments with uncertainty $0.50$ \& $0.75$, than when comparing experiments with uncertainty $1.0$ \& $1.25$. Importantly, this effect is not strong. Hence, we do not expect that it will considerably influence our results.


\subsection{Comparisons of Experiments}
\label{ssec:comparisons_of_experiments}
\begin{figure}[!t]
    \centering
    \subfloat[]{\resizebox{!}{2.8in}{\includegraphics[viewport=0.3cm 0.7cm 6.0cm 17cm]{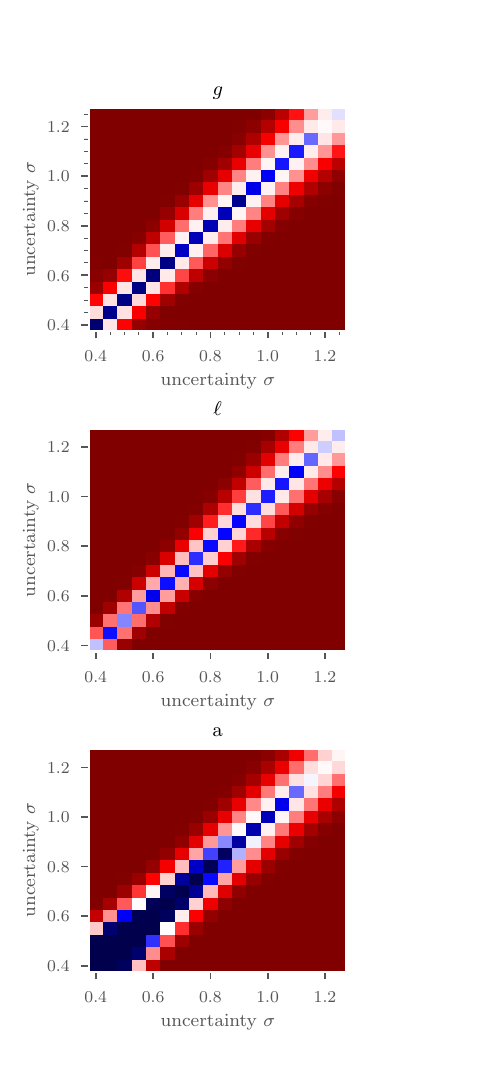}}%
        \label{fig:no_bias_heat_map}}
    \hfil
    \subfloat[]{\resizebox{!}{2.8in}{\includegraphics[viewport=0.3cm 0.7cm 6.0cm 17cm]{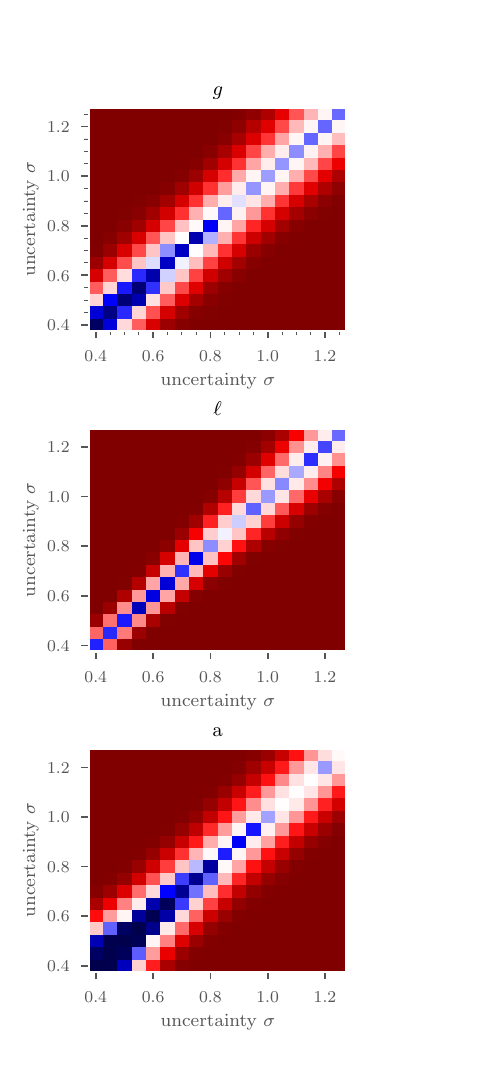}}%
        \label{fig:mixed_bias_heat_map}}
    \hfil
    \subfloat[]{\resizebox{!}{2.8in}{\includegraphics[viewport=0.3cm 0.7cm 8.5cm 17cm]{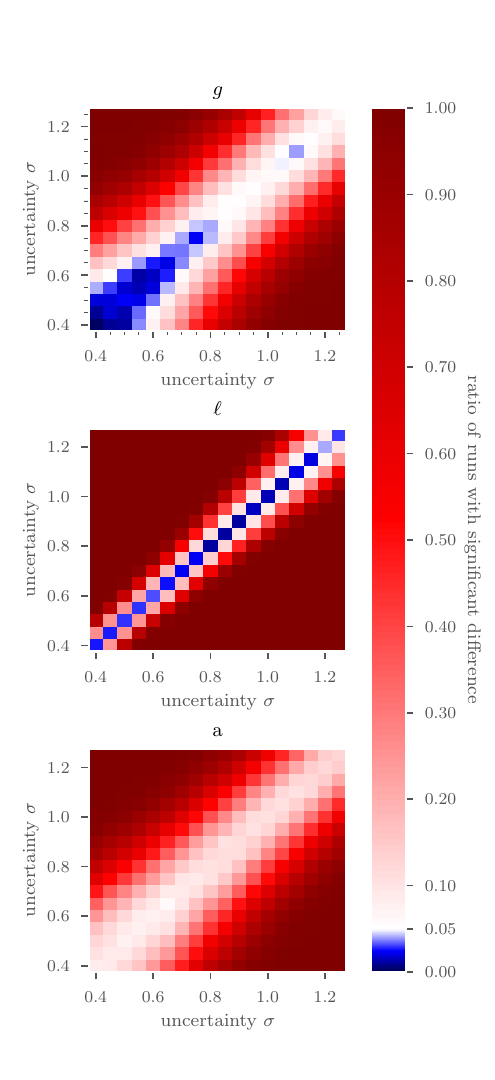}}%
        \label{fig:extreme_bias_heat_map}}
    \caption{The uncertainty vs uncertainty heat maps for the three experiment precision comparison methods. Display (a) presents results for simulated data without subject bias, (b) for simulated data with mixed symmetric bias and (c) for simulated data with extreme symmetric bias.}
    \label{fig:heat_maps}
\end{figure}

\jncomment{I have a feeling we mix too much the description of our methodology with the description of the results. I think it would be good to extend Sec.~\ref{sec:simulation_study_methodology} and leave here only the description of the results.}
To learn how our experiment precision comparison methods perform under a broad range of circumstances, we generated so called \textit{uncertainty vs uncertainty heat maps}. Fig.~\ref{fig:no_bias_heat_map} presents an example of a set of three such heat maps---one per experiment precision comparison method. 

To generate this plot, we take all the simulated experiments (cf. Sec.~\ref{ssec:simulation_parameters_and_scenarios}) and fit to them the models underlying our experiment precision measures. Then, we iterate over all possible uncertainty pairs (e.g., $\{(0.4, 0.4), (0.4, 0.45), (0.4, 0.5), \ldots\}$) and for each pair we randomly choose (with replacement) $2\,000$ simulated experiments. For example, when comparing uncertainty $\sigma=0.4$ with uncertainty $\sigma=0.5$, we sample with replacement $2\,000$ simulated experiments with $\sigma=0.4$ and, likewise, we sample with replacement $2\,000$ simulated experiments with $\sigma=0.5$. To make our approach explicit, let us represent the vector of randomly chosen experiments with $\sigma=0.4$ as $\mathbf{e_{0.4}} = \{e^1_{0.4}, e^2_{0.4}, \ldots, e^{2\,000}_{0.4}\}$ and the vector of randomly chosen experiments with $\sigma=0.5$ as $\mathbf{e_{0.5}} = \{e^1_{0.5}, e^2_{0.5}, \ldots, e^{2\,000}_{0.5}\}$. Now, the pairs we consider in our analysis are as follows: $\{ (e^1_{0.4}, e^1_{0.5}), (e^2_{0.4}, e^2_{0.5}), \ldots\}$.

For each pair of simulated experiments, we apply our experiment precision comparison methods (cf. Sec.~\ref{sec:methods}). Each comparison yields a $p$ value that is significant or not (assuming the 5\% significance level). Our null hypothesis is that there are no statistically significant differences between a pair of experiments. Since we have $2\,000$ simulated experiment pairs per uncertainty-uncertainty pair, we can compute what ratio of comparisons points to a statistically significant difference in experiment precision. This ratio falls within the 0 to 1 range. As there are 18 values of uncertainty $\sigma$ we consider, the complete heat map has $18 \times 18 = 324$ pixels.

Figures \ref{fig:no_bias_heat_map} to \ref{fig:extreme_bias_heat_map} present heat maps for each of the three experiment precision comparison methods. Please note that Fig.~\ref{fig:heat_maps} should be read column by column (there are three heat maps in each column). Fig.~\ref{fig:no_bias_heat_map} presents the results for the simulated experiments without bias. Fig.~\ref{fig:mixed_bias_heat_map} corresponds to the simulated experiments with symmetric mixed bias. Finally, Fig.~\ref{fig:extreme_bias_heat_map} shows the results for the simulated experiments with extreme symmetric bias.

\jncomment{The following paragraph needs revision of other co-authors.}
Ideally, we want an experiment precision method to differentiate between simulated experiments generated with different uncertainty $\sigma$. For example, a simulated experiment with $\sigma=0.4$ should be marked as different, in terms of experiment precision, when compared with an experiment with $\sigma=0.45$. However, two experiments with the same uncertainty, should \textit{not} be flagged as significantly different. Applying these observations to heat maps, we ideally want to see a heat map with ratios of significant differences close to 0.05 on the diagonal (going from the left bottom corner to the top right corner of a heat map) and close to 1 everywhere else. Please note that Figures \ref{fig:no_bias_heat_map} to \ref{fig:extreme_bias_heat_map} use a colour map with dark blue corresponding to the ratio of significant differences of 0, white corresponding to 0.05, and dark red to 1.0.

Looking at Figures \ref{fig:no_bias_heat_map} to \ref{fig:extreme_bias_heat_map}, we see that the $\ell$ measure based method performs best. Its rejection ratios on the diagonal are close to 0.05 and close to 1 everywhere else. Similarly, the method is least affected by changes in the bias level. It also performs equally well in all three bias scenarios. Importantly, the same is \textit{not} true for the other methods. For them, the higher the bias, the worse the performance.

The $g$ and $a$ methods perform worse than the $\ell$ method. Their performance deteriorates as the bias level increases. Moreover, they flag as similar experiments with different uncertainties more often than the $\ell$ method. This is especially true for experiments with similar uncertainties (i.e., for experiment pairs close to the heat map's diagonal).

To quantify the differences between experiment precision comparison methods, we measure the distance between each method's heat map and the ideal heat map (i.e., the one with rejection ratios equal to $0.05$ on the diagonal and $1$ everywhere else). We compute this distance using the Mean Absolute Error (MAE) measure. Tab.~\ref{tab:mae_results} presents the results. Again, the $\ell$ method performs best in all bias scenarios. It is clearly visible that the method is robust to bias level changes. The $g$ method is second best in the no bias scenario, with the $a$ method following closely. In the mixed symmetric bias scenario, the performance of the $g$ and $a$ methods is similar, although it is worse than in the no bias scenario. Finally, in the extreme symmetric bias scenario, only the $\ell$ method performs reasonably well. (Please keep in mind that the rejection ratio spans the range of 0 to 1. Thus, the MAE close to 0.5 corresponds to being, on average, half the scale range away from the desired result.)
\begin{table}[!t]
    \centering
    \begin{threeparttable}
    \caption{Distances Between Method's Heat Map and the Ideal Heat Map for Each Bias Scenario.}
    \label{tab:mae_results}
    \begin{tabular}{@{}p{1.2cm}p{1.7cm}p{1.7cm}p{1.7cm}@{}}
    \toprule
    \multirow{2}{*}{Method} & \multicolumn{3}{l}{MAE~$\downarrow$}                             \\ \cmidrule(l){2-4} 
                            & No Bias         & Mixed Bias      & Extreme Bias    \\ \cmidrule(r){1-1}
    $\ell$                  & \textbf{0.1537} & \textbf{0.1533} & \textbf{0.1611} \\
    $g$                     & 0.2028          & 0.2810          & 0.4790          \\
    $a$                     & 0.2397          & 0.2749          & 0.4099          \\ \bottomrule
    \end{tabular}
    \begin{tablenotes}
        \item MAE stands for mean absolute error. The least distance for each bias scenario is given in bold.
    \end{tablenotes}
    \end{threeparttable}
\end{table}


\section{Practical Use Case}
\label{sec:practical_use_case}
\jntodo{Consider explaining here that the content of this section is divided into two parts: i) a comparison of precision measures performance for QoE experiments of three types (VR, speech, and video) and ii) an example how precision measures properly detect the lack of precision of VIME1 and CCRIQ2 experiments.}
Here, we present how the experiment precision measures perform in practice. This practical example is an indication of what can be achieved using the experiment precision measures. We show that our experiment precision measures are able to show differences between different types of subjective experiments. Importantly, measures' indications are in line with our expectations regarding experiment type's precision. We hope that this section will convince the reader of the practical value our notion of experiment precision brings.

Our analysis is based on subjective experiments of three types: i) virtual reality (VR) QoE, ii) speech QoE, and iii) video QoE. Specifically, we use one VR QoE experiment, one speech QoE experiment, and six video QoE experiments. The VR experiment comes from an international multilaboratory QoE study~\cite{Gutierrez2021}. The one experiment we use from this study occurred in Wuhan, used the ACR methodology, and made use of video sequences being 10 s long. (From now on we refer to this experiment as VR.) The speech QoE experiment comes from the ``ITU-T Coded-Speech Database'' (constituting Supplement 23 to the P series of ITU-T Recommendations)~\cite{itutsupp23}. We use only the responses collected by Nortel during the first of the three experiments reported in this database. (We later refer to this experiment as S.) Finally, the six video QoE experiments we use come from the international multilaboratory VQEG HDTV Phase I study~\cite{HDTV_Phase_I_test}. (We refer to the experiments in this study as V-$n$, with $n \in \{1, 2, \ldots, 6\}$.)

Following our intuition and SOS--MOS curves for various types of subjective experiments (cf. Fig.~5 in~\cite{hosseld2011sos}), we expect video and speech QoE experiments to be \textit{more} precise than VR QoE experiments. If we were to take the SOS--MOS curves presented in~\cite{hosseld2011sos} for granted, we should also expect video QoE experiments to be \textit{more} precise than speech QoE experiments.
(Speech QoE experiments correspond to VoIP experiments in Fig.~5 from~\cite{hosseld2011sos}.)
All in all, our expected ordering of experiment types in terms of precision (starting from the least precise experiment type) is as follows: VR, speech, and video.

\begin{table}[!t]
    \centering
    \begin{threeparttable}
    \sisetup{ 
         input-decimal-markers = {.}, 
         group-separator={}, 
         table-text-alignment = center}
    \caption{Experiment Precision Measures For QoE Subjective Experiments of 3 Types: VR(VR), Speech (S), and Video (V-$n$).}
	\label{tab:raw_exp_prec}
	\begin{tabular}{p{.9cm}*{6}{S[table-format=1.4]}}
     \toprule
       {Exp.}  & {$\ell$~$\downarrow$} & {SE($\ell$)} & {$g$~$\uparrow$} & {SE($g$)} & {$a$~$\downarrow$} & {SE($a$)}\\
     \midrule
    V-6 & 0.574 & 0.014 & 0.908 & 0.0050 & 0.137 & 0.0020 \\
    V-1 & 0.583 & 0.011 & 0.891 & 0.0068 & 0.149 & 0.0022 \\
    V-4 & 0.610 & 0.020 & 0.826 & 0.0056 & 0.224 & 0.0021 \\
    V-3 & 0.613 & 0.016 & 0.863 & 0.0066 & 0.188 & 0.0021 \\
    V-5 & 0.627 & 0.019 & 0.871 & 0.0059 & 0.190 & 0.0021 \\
    V-2 & 0.627 & 0.022 & 0.867 & 0.0070 & 0.191 & 0.0021 \\
    S   & 0.953 & 0.028 & 0.744 & 0.0083 & 0.281 & 0.0015 \\
    VR  & 1.059 & 0.037 & 0.692 & 0.0093 & 0.335 & 0.0040 \\
     \bottomrule
    \end{tabular}
    \begin{tablenotes}
        \item Arrows point in the direction of high precision. SE$(\cdot)$ stands for standard error of a particular precision measure.
    \end{tablenotes}
    \end{threeparttable}
\end{table}

Table~\ref{tab:raw_exp_prec} presents experiment precision results for eight subjective experiments of interest. The table is sorted in ascending order, according to the measure $\ell$. The arrows next to the headings identifying three experiment precision measures indicate the direction of higher precision. For example, the higher the $g$, the more precise the experiment is. Our experiment precision measures order three experiment types in line with the prior expectations. That is, the one VR experiment is assessed to be the least precise, with the speech experiment following, and the six video experiments assessed to be the most precise. Importantly, this ordering is reflected in the readings from all three experiment precision measures. This suggests that all measures are capable of estimating subjective experiment precision in line with the intuition of experts.

In practice, we may need to compare a pair of subjective experiments in terms of their precision. Thus, it is interesting to check whether our experiment precision measures indicate statistically significant differences between each pair of experiments from our pool of eight experiments. We generally expect to see statistically significant differences between experiments of different types (e.g., VR QoE vs speech QoE). We do not expect statistically significant differences between experiments of the same type (e.g., V-1 experiment vs V-2 experiment).
\begin{table}[!t]
    \centering
    \begin{threeparttable}
    \caption{$p$-Values Resulting From Comparing Experiment Precision Measures Between All Video QoE Experiments. We Expect No Statistically Significant Differences.
    \jntodo{If time permits, change the notation style to the \LaTeX~mathematical mode.}}
	\label{tab:same_type_pair_comparisons}
	 \begin{tabular}{p{1.0cm}p{1.0cm}*{3}{p{1.3cm}}}
     \toprule
       {Exp. 1}  & {Exp. 2} & {$\ell$ $p$-Value} & {$g$ $p$-Value} & {$a$ $p$-Value} \\
     \midrule
    V-6 & V-1 & \cellcolor{blue!0}6.07E-01 & \cellcolor{blue!25}4.76E-02 & \cellcolor{blue!25}4.44E-05 \\
    V-6 & V-4 & \cellcolor{blue!0}1.45E-01 & \cellcolor{blue!25}5.76E-24 & \cellcolor{blue!25}2.60E-97 \\
    V-6 & V-3 & \cellcolor{blue!0}7.70E-02 & \cellcolor{blue!25}1.21E-07 & \cellcolor{blue!25}2.11E-50\\
    V-6 & V-5 & \cellcolor{blue!25}2.97E-02 & \cellcolor{blue!25}3.10E-06 & \cellcolor{blue!25}9.07E-53\\
    V-6 & V-2 & \cellcolor{blue!0}5.10E-02 & \cellcolor{blue!25}3.72E-06 & \cellcolor{blue!25}5.10E-53\\
    V-1 & V-4 & \cellcolor{blue!0}2.36E-01 & \cellcolor{blue!25}1.23E-12 & \cellcolor{blue!25}1.21E-78\\
    V-1 & V-3 & \cellcolor{blue!0}1.30E-01 & \cellcolor{blue!25}3.12E-03 & \cellcolor{blue!25}4.60E-32\\
    V-1 & V-5 & \cellcolor{blue!25}4.89E-02 & \cellcolor{blue!25}2.70E-02 & \cellcolor{blue!25}2.30E-34\\
    V-1 & V-2 & \cellcolor{blue!0}8.29E-02 & \cellcolor{blue!25}1.57E-02 & \cellcolor{blue!25}1.35E-34\\
    V-4 & V-3 & \cellcolor{blue!0}9.12E-01 & \cellcolor{blue!25}3.21E-05 & \cellcolor{blue!25}2.06E-28\\
    V-4 & V-5 & \cellcolor{blue!0}5.42E-01 & \cellcolor{blue!25}7.03E-08 & \cellcolor{blue!25}1.03E-25\\
    V-4 & V-2 & \cellcolor{blue!0}5.73E-01 & \cellcolor{blue!25}5.33E-06 & \cellcolor{blue!25}1.71E-25\\
    V-3 & V-5 & \cellcolor{blue!0}5.77E-01 & \cellcolor{blue!0}3.54E-01 & \cellcolor{blue!0}4.97E-01\\
    V-3 & V-2 & \cellcolor{blue!0}6.11E-01 & \cellcolor{blue!0}6.27E-01 & 4.59E-01\\
    V-5 & V-2 & \cellcolor{blue!0}9.96E-01 & \cellcolor{blue!0}6.95E-01 & \cellcolor{blue!0}9.51E-01\\
    \bottomrule
    \end{tabular}
    \begin{tablenotes}
    \item $p$-Values smaller than 0.05 are marked with purple background.
    \end{tablenotes}
    \end{threeparttable}
\end{table}

\begin{table}[!t]
    \centering
    \begin{threeparttable}
    \caption{$p$-Values Resulting From Cross-type Comparisons of Experiment Precision Measures. We Expect Statistically Significant Differences.
    \jntodo{If time permits, change the notation style to the \LaTeX~mathematical mode.}}
	\label{tab:cross_type_pair_comparisons}
	\begin{tabular}{p{1.0cm}p{1.0cm}*{3}{p{1.3cm}}}
     \toprule
       {Exp. 1}  & {Exp. 2} & {$\ell$ $p$-Value} & {$g$ $p$-Value} & {$a$ $p$-Value} \\
     \midrule
    V-6 & S  & \cellcolor{blue!25}6.02E-14 & \cellcolor{blue!25}3.29E-45 & \cellcolor{blue!25}1.99E-167 \\
    V-1 & S  & \cellcolor{blue!25}3.09E-13 & \cellcolor{blue!25}1.84E-34 & \cellcolor{blue!25}9.65E-148 \\
    V-4 & S  & \cellcolor{blue!25}1.21E-12 & \cellcolor{blue!25}3.82E-15 & \cellcolor{blue!25}7.02E-66\\
    V-3 & S  & \cellcolor{blue!25}1.07E-12 & \cellcolor{blue!25}5.82E-25 & \cellcolor{blue!25}1.34E-112\\
    V-5 & S  & \cellcolor{blue!25}4.28E-12 & \cellcolor{blue!25}2.06E-29 & \cellcolor{blue!25}3.14E-109\\
    V-2 & S  & \cellcolor{blue!25}1.06E-11 & \cellcolor{blue!25}7.78E-26 & \cellcolor{blue!25}5.03E-109\\
    V-6 & VR & \cellcolor{blue!25}1.10E-14 & \cellcolor{blue!25}1.49E-37 & \cellcolor{blue!25}3.78E-66\\
    V-1 & VR & \cellcolor{blue!25}3.95E-14 & \cellcolor{blue!25}9.45E-36 & \cellcolor{blue!25}3.49E-65\\
    V-4 & VR & \cellcolor{blue!25}8.09E-14 & \cellcolor{blue!25}1.46E-22 & \cellcolor{blue!25}3.33E-44\\
    V-3 & VR & \cellcolor{blue!25}1.09E-13 & \cellcolor{blue!25}5.90E-30 & \cellcolor{blue!25}8.57E-55\\
    V-5 & VR & \cellcolor{blue!25}2.62E-13 & \cellcolor{blue!25}1.20E-31 & \cellcolor{blue!25}2.03E-54\\
    V-2 & VR & \cellcolor{blue!25}3.41E-13 & \cellcolor{blue!25}7.54E-31 & \cellcolor{blue!25}2.28E-54\\
    S & VR   & \cellcolor{blue!25}2.54E-02 & \cellcolor{blue!25}5.42E-05 & \cellcolor{blue!25}2.52E-21\\
    \bottomrule
    \end{tabular}
    \end{threeparttable}
\end{table}

Tables~\ref{tab:same_type_pair_comparisons} and \ref{tab:cross_type_pair_comparisons} show $p$-values resulting from comparing experiment precision measures between all pairs of the eight experiments of interest. The headings identify which column corresponds to which experiment precision measure. The first two columns indicate which two experiments are compared. We highlight in purple comparisons that resulted in significant differences (assuming a 5\% significance level).
Note that Tab.~\ref{tab:same_type_pair_comparisons} presents same-type pairs (i.e., the two experiments in the pair are of the same type), whereas Tab.~\ref{tab:cross_type_pair_comparisons} presents cross-type pairs (i.e., the two experiments in the pair are of two different types).
All precision measures flag cross-type pairs as corresponding to significant differences in experiment precision. This is desirable. However, out of 15 same-type pairs (cf. Tab.~\ref{tab:same_type_pair_comparisons}), the measures $g$ and $a$, mark 12 as indicating significant differences in terms of precision. This is counter-intuitive and suggests that there may be a problem with these two measures. The behaviour of the $\ell$ measure is generally in line with our prior expectations. It flags as significantly different only two out of 15 same-type pairs. It is worth keeping in mind that since we assume the 5\% significance level, flagging roughly one comparison as significant may happen purely due to randomness. Thus, the measure flagging as significant only two out of 15 same-type pairs is very close to our prior expectation of no true differences.

\subsection{Detecting Imprecise Experiments}
\label{ssec:detecting_imprecise_experiments}
It is interesting to check whether the precision measures would be able to detect problems with experiments that are known to be flawed. One such example of experiments with insufficient precision are experiments VIME1 and CCRIQ2 described in~\cite{Nawala2020}. (We later refer to these experiments as V\&C.) The two experiments investigated the quality of a set of images taken with consumer capture devices (e.g., smartphones or tablets). Due to a few unusual experiment design choices, the subjective responses gathered during the two experiments were identified in \cite{Nawala2020} as less precise than would be typical for a standard image QoE subjective experiment.

To check whether the experiment precision measures are able to detect the low precision of V\&C, we apply the measures to raw subjective responses. In other words, we do \emph{not} apply any data cleansing procedure before running the measures. The only preprocessing step that we take is to remove the responses of two subjects---one with ID 259 from the VIME1 experiment and one with ID 270 from the CCRIQ2 experiment. We do so since the two subjects did not assess the quality of all stimuli. The current implementation of our precision measures assumes that all subjects assess the quality of all stimuli. Thus, we had to discard the responses of subjects for whom this was not the case.

Table~\ref{tab:ccriq2_vime1_prec_meas} presents the results of applying our measures to the data originating from V\&C experiments.
The first thing to notice is how the results compare to the results presented in Tab.~\ref{tab:raw_exp_prec}. According to the precision measures, experiments V\&C have the precision similar to the VR QoE experiment.
Both VIME1 and CCRIQ2 are also statistically significantly \emph{less} precise than the speech QoE experiment and this is true for all precision measures (assuming a 5\% significance level). This is unusual. According to \cite{hosseld2011sos}, image QoE experiments are generally \emph{more} precise than speech QoE experiments (cf. Fig.~5 of \cite{hosseld2011sos}). Precision smaller than that of the speech QoE experiment and similar to the VR QoE experiment indicates that there is a problem with the precision of V\&C experiments.
\begin{table}[!t]
    \caption{Raw Experiment Precision Measures Results For Two Image QoE Experiments---VIME1 (I-V) and CCRIQ2 (I-C).}
	\label{tab:ccriq2_vime1_prec_meas}
	\centering
     \begin{tabular}{l*{6}{S[table-format=1.4]}}
     \toprule
       {Exp.}  & {$\ell$~$\downarrow$} & {SE($\ell$)} & {$g$~$\uparrow$} & {SE($g$)} & {$a$~$\downarrow$} & {SE($a$)}\\
     \midrule
    I-V & 1.053 & 0.0330 & 0.717 & 0.0085 & 0.314 & 0.0025 \\
    I-C & 1.100 & 0.0316 & 0.683 & 0.0103 & 0.347 & 0.0030 \\
     \bottomrule
     \end{tabular}
\end{table}

As image QoE experiments, V\&C should have the precision similar to that of other image or video QoE experiments~\cite{hosseld2011sos}. Table~1 in \cite{hosseld2011sos} shows that typical image and video experiments correspond to $a$ between $0.0377$ and $0.2116$. However, VIME1 and CCRIQ2 correspond to $a$ of $0.314$ and $0.347$, respectively. Such high readings of $a$ make the two experiments resemble cloud gaming QoE experiments. Significantly, Hoßfeld \textit{et al.} classify cloud gaming experiments as one of the \emph{least} precise QoE experiments. Taking this and previous observations into account, it is clear that our experiment precision measures correctly detected the low precision of V\&C experiments.


\section{Discussion}
\label{sec:discussion}
The results in Sec.~\ref{sec:practical_use_case} suggest that the $\ell$ experiment precision measure is the most reliable one. Our intuition here is that this measure's dominance over the two other measures stems from its ability to ignore subject bias. There are at least two examples that confirm this ability. The first is given in Fig.~\ref{fig:upsilon_bar_vs_no_bias_prob}. There, it is clearly visible that the $\ell$ measure is almost constant across various bias probabilities. The second example of $\ell$ measure's ability to ignore subject bias is given in Tab.~\ref{tab:same_type_pair_comparisons}. From Tab.~1 in \cite{Janowski2014}, we know that the HDTV4 experiment (denoted V-4 in Tab.~\ref{tab:same_type_pair_comparisons}) recruited people with subject biases significantly exceeding (in terms of its range of values) subject biases present in all other HDTV experiments. Yet, the $\ell$ measure does not flag this experiment as significantly different from other HDTV experiments. This is a strong suggestion that this measure follows our theoretical assumptions regarding the notion of experiment precision (cf. Sec.~\ref{ssec:notion_of_experiment_precision}) and truly compares experiments without taking into account potential differences in response biases.

We stated in Sec.~\ref{sec:practical_use_case} that both the $g$ and $a$ experiment precision measures labelled 12 out of 15 same-type experiment pairs as significantly different. The same is not true for the $\ell$ measure, which detected significant differences in only two pairs. Although these statements point to $\ell$ measure's superiority, there is one caveat that we must mention. The $g$ and $a$ measures internally use per stimulus estimated parameters. For the case of HDTV experiments, this means that the two measures operate on 168 stimuli (i.e., on a sample with 168 observations). On the other hand, the $\ell$ measure internally uses per subject estimated parameters. Since there were 24 participants in each HDTV experiment, the $\ell$ measure operates on a sample of 24 observations. Now, in general, the more observations in a sample, the more precise the estimation process. It is thus natural that $g$ and $a$ measures are more sensitive to changes in experiment precision than the $\ell$ measure. Hence, there is a chance that the increased sensitivity of $g$ and $a$ measures is responsible for the higher number of significant differences detected by these measures. In other words, it is possible that $\ell$ measure's seeming superiority stems from its lesser sensitivity to experiment precision changes, rather than from its true accuracy.

There is another caveat related with the $\ell$ measure (and the appropriate experiment precision comparison method). Although the results in Sec.~\ref{ssec:comparisons_of_experiments} point to $\ell$ measure's superiority, the measure has a property that may give it an unfair advantage. Internally, the $\ell$ measure is based on the Li2020 model. This model is built on top of the normal distribution, the same distribution our data generator uses to generate simulated experiments. Thus, it may be easier for the $\ell$ measure, when compared to other measures, to properly recover uncertainty $\sigma$ that we use to generate the simulated data. Since we assume that differences in data generator's $\sigma$ correspond to changes in experiment precision, a measure better able to recover $\sigma$ naturally appears as better in detecting experiment precision differences. Importantly, the superiority in recovering $\sigma$ of the data generator does not have to necessarily translate into proper experiment precision extraction for real data.

Notwithstanding the caveats mentioned above, the $\ell$ measure seems to be the best in estimating experiment precision (at least within the boundaries of our definition of the concept given in Sec.~\ref{ssec:notion_of_experiment_precision}). Its superior performance is clear both in the simulated experiments (cf. Sec.~\ref{sec:results}) and when the measure is applied to real data (cf. Sec.~\ref{sec:practical_use_case}). Still, our recommendation is to compute all three measures ($\ell$, $g$, and $a$) for each subjective experiment conducted. Measure's indications should be reported, along with their standard errors and the number of observations they are based on. For example, if we were to report the $g$ measure for the sixth experiment of the VQEG HDTV Phase I study (cf. the first row of Tab.~\ref{tab:raw_exp_prec}), we would give measure's indication ($0.908$), its standard error ($0.0050$) and the number of observations it is based on ($168$, since that many video stimuli were presented to experiment participants).

When researching the topic of experiment precision, we came up with yet another experiment precision comparison method. It is based on the ratio of variance for similar MOS regions for a pair of experiments. Although conceptually promising, the method failed to provide satisfactory performance. We refer readers interested in learning more about this method to Appendix~\ref{app:exp_prec_comp_paired_var} in the Supplemental material.

\ljtodo{Read the following paragraph and assess whether it convincingly says why we do \textit{not} use a measure based on total variance of subjective responses.}
Considering our definition of experiment precision, one may wonder why we do not use a measure as simple as taking the total variance of subjective responses. In other words, why not put all responses in a single large vector and calculate its variance. The answer here is that we know that, in principle, such a measure would be quickly fooled by subject bias. In our simulation study, we introduce subject bias that shifts all responses of a single subject. Since we implement this shift as a change in data generator's true average user rating $\mu_x$ (cf. (\ref{eq:data_generator})), we end up with a normal distribution underlying our data generator with a significant probability mass to the left of 1 or to the right of 5. Taking into account that our data generator censors the domain of the underlying normal distribution to the range 1 to 5, the variance of responses generated under these conditions has a highly nonlinear relationship to the data generator's uncertainty parameter $\sigma_u$. Since we assume that $\sigma_u$ corresponds to experiment precision, relying only on the variance of responses generated, we have a very small chance of properly capturing experiment precision.

It may be surprising for some readers that the standard error of the measure $a$ for the speech QoE experiment is smaller than that of the video QoE experiments we investigated (cf. Tab.~\ref{tab:raw_exp_prec}). Since video experiments were labeled by our measures as more precise than the speech experiment, it is tempting to assume that the standard error for any precision measure should be smaller for these experiments as well. To understand why this does not need to be the case for the measure $a$, we refer the reader to Fig.~\ref{fig:ghost_like_shape} in the Supplemental material. The figure shows that a set of possible MOS--variance pairs, and likewise the MOS--SOS pairs, is bounded by a ghost-like shape. For example, a set of stimuli with minimum possible variance would correspond to MOS--variance pairs falling along a wave-like pattern at the bottom of the shaded area in Fig.~\ref{fig:ghost_like_shape}. If we tried to fit a quadratic function to this shape (as the measure $a$ internally does), the resultant fit would be rather poor. Hence, it is possible and justified that a set of highly precise subjective responses corresponds to the measure $a$ that has a high standard error. Furthermore, it is natural that the fit of the quadratic function improves when we move away from the bottom of the shaded area in Fig.~\ref{fig:ghost_like_shape}.

\jncomment{I would really like others to take a look at the following paragraph.}
Although we present a set of measures assessing experiment precision, we would like to stress that these measures are \textit{not} sufficient to compare the precision of a pair of experiments. In other words, our measures should not be used as the \emph{only} mean used to compare a pair of experiments in terms of their precision. Experiment precision is a multifaceted concept. We thus recommend to approach the topic comprehensively. When comparing a pair of subjective experiments in terms of their precision, we suggest to consider, among others, the following factors: i) inter-rater reliability, ii) the number of subjects discarded due to the post-experimental screening of subjects using Pearson linear correlation (cf. clause 11.4 of \cite{P913_2021}), iii) width of MOS confidence intervals (cf. clause A1-2.2 of \cite{BT500}), and iv) to what extent an experiment conforms to the ITU and the research community guidelines.


\section{Conclusions and Further Work}
\label{sec:conclusions_further_work}
In this work, we propose a notion of experiment precision. We also define and test three experiment precision measures and related experiment precision comparison methods. We do so both through a simulation study and by using real data from subjective experiments on VR, speech, image, and video QoE. We provide guidelines regarding reporting experiment precision measures as well. We hope these guidelines will be followed by practitioners creating new subjective data sets. At last, our results suggest that the Li2020 model~\cite{Li2020simple} based $\ell$ measure performs best in assessing experiment precision.

The content of this paper provides answers to the research questions we posed in Sec.~\ref{ssec:research_questions}. If it comes to research question R1, we show that all three measures we put forward (the $g$, $a$, and $\ell$ measure) properly represent experiment precision of both simulated and real subjective MQA experiments (cf. Sec.~\ref{ssec:systematic_analysis} and Sec.~\ref{sec:practical_use_case}). Naturally, the caveat here is that this is true within the confines of our definition of experiment precision. We also demonstrate that only the $\ell$ measure is able to capture (and then discard) subject bias. With respect to research question R2, we propose three methods to statistically compare precision of two subjective experiments (cf. Sec.~\ref{sec:methods} for their definition and Sec.~\ref{ssec:comparisons_of_experiments} for their performance review). In Sec.~\ref{ssec:comparisons_of_experiments} we illustrate how our experiment comparison methods behave in different scenarios. This addresses research question R3. All three methods properly capture response uncertainty. However, only the $\ell$ method can deal with (even significant) subject bias well. If it comes to research question R4, in Sec.~\ref{sec:practical_use_case} we demonstrate that all three measures order (in terms of experiment precision) the three types of real-world subjective experiments (VR, speech, and video QoE) identically. Still, the experiment precision ordering of individual experiments within one experiment type differs depending on which measure we use. Nonetheless, the indications of all three measures are in line with expert intuition and domain knowledge regarding the three experiment types investigated.

We believe that with this work we provide sufficient evidence to support the claims we put forward in Sec.~\ref{ssec:claims_contributions}. Specifically, our experiment precision measures prove to be able to position an experiment in terms of its precision in relation to various experiment types (e.g., video or speech QoE experiments). Our experiment precision comparison methods make it possible to statistically compare (in terms of precision) multiple experiment runs. Finally, the results in Sec.~\ref{sec:results} and \ref{sec:practical_use_case} make it clear that the $\ell$ measure should be a go-to measure for assessing experiment precision.

If future work is concerned, we consider repeating the simulation study using a different data generator. Specifically, we would like to run the study using a data generator based on the beta distribution. Our hope is that this would eliminate the unfair advantage the $\ell$ measure has over the other measures when the normal distribution based data generator is used. The other direction of future research relates to creating a single experiment precision measure. We predict that through machine learning techniques we should be able to propose one main experiment precision measure. It would internally use the results of the three existing measures. Hopefully, based on the three already defined measures, it would be able to assess experiment precision better than any single measure.

We hope that our notion of experiment precision will help MQA practitioners differentiate between subjective experiments. In particular, we envision that experiment precision measures may be part of a set of tools aimed at detecting differences between subjective experiments performed following distinct methodologies. For example, the notion of experiment precision may help decide which experiment methodology results in generally more precise responses. The two methodologies compared may, for example, use two different response scales (a five-point scale vs seven-point scale). Thanks to our experiment precision measures, it may be easier to decide which experiment methodology should be followed when response precision is key.

To facilitate adoption of our experiment precision measures, we make available a source code implementing the experiment precision measures and experiment comparison methods presented in this paper. The source code is available under the following link: \url{https://github.com/Qub3k/qoe-experiment-precision}.



%





\ifCLASSOPTIONcaptionsoff
  \newpage
\fi



%
\bibliographystyle{IEEEtran}
\bibliography{main}

\cleardoublepage  


\appendices

\section*{Supplemental Material}

\section{Comparison of Two \textit{a} Measures With the \textit{t}-Test}
\label{app:comp_of_two_a_measures}
The basic idea is to compute the SOS parameters $a_1$ and $a_2$ for the two experiments being compared. Having done that, we can perform a two-sample independent $t$-test to check whether the parameter estimates are statistically significantly different.
Formally, we consider two experiments $i\in\{1,2\}$, which consist of $K_i$ test stimuli each (e.g., $K_i=\{168, 144\}$ videos).
%
%
%
For experiment $i$, the MOS values $m_{i,x}$ and variances $v_{i,x}$ yield the SOS parameter $a_i$ according to (\ref{eq:sos}) (with $x \in (1, 2, \ldots, K_i)$). Specifically, we need to estimate $a_i$ for both experiments from the pair using the following formula:
\begin{align}
 & a_i = \frac{\sum_{x=1}^{K_i} (5-m_{i,x}) \cdot (m_{i,x}-1) \cdot v_{i,x}}{\sum_{x=1}^{K_i} (5-m_{i,x})^2 \cdot (m_{i,x}-1)^2}.
\end{align}

Now, we can use a standard $t$-test (Welch's unequal variances $t$-test) to compare if $a_1$ and $a_2$ are statistically significantly different. The $t$-test takes into account: i) the parameters $a_1, a_2$ estimated from the MOS--SOS tuples, ii) variance $\nu_1, \nu_2$ of the parameter estimates (from the quadratic OLS regression), and iii) $K_1$ and $K_2$.
The unbiased variance of the SOS parameter estimator is
\begin{align}
& \nu_i = \left({\sum_{x=1}^{K_i} (5-m_{i,x})^2 \cdot (m_{i,x}-1)^2} \right)^{-1} \;. 
\end{align}
The $t$ statistic for unequal variances and unequal sample sizes is
\begin{align}
&t_{stat} = \frac{a_1 - a_2}{\sqrt{\frac{\nu_1}{K_1}+\frac{\nu_2}{K_2}}} 
\intertext{with the degrees of freedom}
& d.f. = \frac{ \left({\nu_1}/{K_1} + {\nu_2}/{K_2} \right)^2}{ \frac{(\nu_1/K_1)^2}{K_1-1} +\frac{(\nu_2/K_2)^2}{K_2-1} } \,.
\end{align}
Finally, we can compute the $p$-value of the two-sided test. Since the test statistic comes from the Student's $t$-distribution (with its cumulative distribution function denoted as $F_T(t_{stat}, d.f.)$), the $p$-value is
\begin{align}
p = 2 \cdot (1 - F_T(t_{stat}, d.f)).
\end{align}
If $p> \alpha$ (where $\alpha$ is the significance level --- 0.05 in most cases), then the null hypothesis is accepted. Importantly, the null hypothesis is that the two SOS parameters $a_1$ and $a_2$ are equal. Otherwise (i.e., when $p\leq \alpha)$, the null hypothesis is rejected. This corresponds to concluding that the parameters $a_1$ and $a_2$ are statistically significantly different.

Importantly, since we assume that SOS parameter $a$ is equal in value to the experiment precision measure $a$. Conclusions reached for a pair of SOS parameters $a$ are equally applicable to a pair of $a$ experiment precision measures.


\section{Experiment Precision Comparison Method Based on Paired Variances}
\label{app:exp_prec_comp_paired_var}
This approach compares a pair of experiments in terms of response variance for similar MOS regions. These regions correspond to stimuli having similar MOS values. In other words, conventionally, one MOS value corresponds to a single stimulus. To make sure we estimate response variance with high precision, we combine the responses assigned to stimuli with similar MOS values. This means that responses assigned to more than one stimulus may be part of a single MOS region. Specifically, we combine responses assigned to stimuli with MOS difference not greater than 0.2. For example, let us compare Experiment 1 with MOS values $\{2.0, 3.1, 4.0, 5.0\}$ and Experiment 2 with MOS values $\{2.2, 3.0, 4.5, 5.0\}$. Now, following our method, the responses assigned to the stimulus with MOS equal 2.0 (Experiment 1) and the responses assigned to the stimulus with MOS 2.2 (Experiment 2) are considered to belong to the same MOS region, centred at $\text{MOS}=2.1$. Similarly, the responses at MOS 3.1 (Experiment 1) and at MOS 3.0 (Experiment 2) are considered to belong to the merged MOS region centred at 3.05.

The careful reader will notice that there may be MOS regions, which have only been evaluated in one experiment. Considering again the example above, there is a stimulus in Experiment 1, which corresponds to MOS 4.0, but there are no stimuli in Experiment 2 in the same MOS region. Similarly, the MOS region centred at 4.5 is only covered in Experiment 2, but not in Experiment 1. To overcome this issue, we leverage the SOS hypothesis \cite{hosseld2011sos}. For each experiment, we take the MOSs and response variances of all stimuli and fit the corresponding MOS--SOS curve (cf. (\ref{eq:mossos})). As the fitted MOS--SOS curve allows to approximate expected response variances for all MOS regions, we can compute the missing response variances for MOS regions which have not been evaluated in the experiment of interest (such as $SOS^2_2(4.0)$ and $SOS^2_1(4.5)$ in the example above, where $SOS_i(y)$ stands for the standard deviation of scores in the $i$-th experiment for the $\text{MOS}=y$).

Having a paired data set of response variances for the corresponding MOS regions in the two experiments being compared, we move on to formally testing for differences between the two experiments. Since, to the best of our knowledge, there is no statistical method to compare a set of paired variances on a global level (i.e., family-wise), we resort to testing each pair of variances individually. We compare each pair using a modified version of the $F$-test. This results in a vector of $p$-values. Based on this, after applying the relevant correction to account for multiple comparisons, we arrive at a single $p$-value. Comparing this $p$-value to the assumed 5\% significance level, we conclude whether a pair of experiments differ significantly in terms of experiment precision.

For a detailed description of our approach to hypothesis testing for a data set of paired variances, we refer the reader to Appendix~\ref{app:comp_pair_var_w_f_test} in the Supplemental material.

\begin{figure}[!t]
    \centering
    \subfloat[]{\resizebox{!}{.975in}{\includegraphics[viewport=0cm 0cm 6.2cm 6cm]{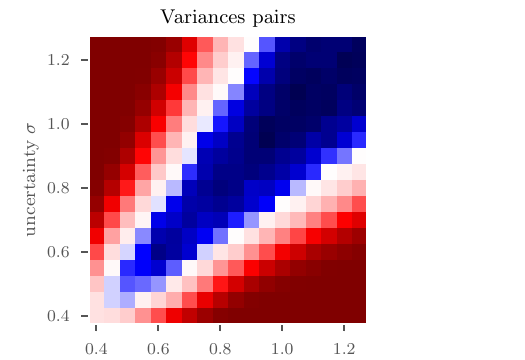}}%
        \label{fig:michi_no_bias_heat_map}}
    \hfil
    \subfloat[]{\resizebox{!}{.975in}{\includegraphics[viewport=0cm 0cm 6.2cm 6cm]{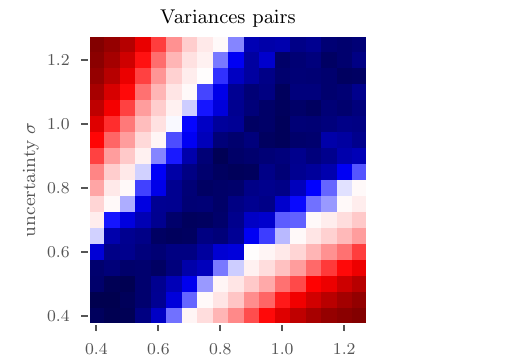}}%
        \label{fig:michi_mixed_bias_heat_map}}
    \hfil
    \subfloat[]{\resizebox{!}{.975in}{\includegraphics[viewport=0cm 0cm 8.6cm 6cm]{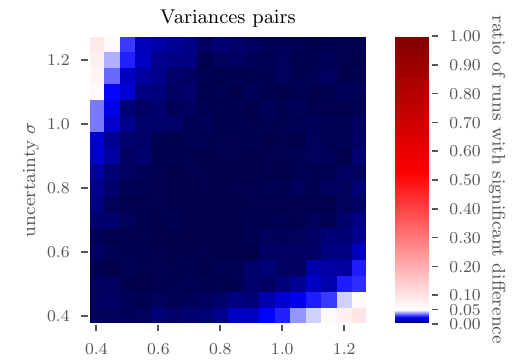}}%
        \label{fig:michi_extreme_bias_heat_map}}
    \caption{The uncertainty vs uncertainty heat maps for the paired variances experiment precision comparison method. Display (a) presents results for simulated data without subject bias. Display (b) presents results for simulated data with mixed symmetric bias and display (c) presents results for simulated data with extreme symmetric bias.}
    \label{fig:michi_heat_maps}
\end{figure}

Following the procedure from Sec.~\ref{ssec:comparisons_of_experiments}, we test how the paired variances experiment precision comparison method performs under various conditions. Figure~\ref{fig:michi_heat_maps} presents the results of this analysis. By comparing this figure to Fig.~\ref{fig:heat_maps}, we see that the performance of the paired variances method is significantly worse than that of the other methods. If it comes to the distance between each heat map in Fig.~\ref{fig:michi_heat_maps} and the ideal heat map (in terms of mean absolute error; cf. Tab.~\ref{tab:mae_results}), the results are as follows: $0.6006$ for the no bias scenario, $0.8098$ for the mixed symmetric bias, and $0.9405$ for the extreme symmetric bias. Again, these results are significantly worse than the results for other methods.


\section{Comparing Paired Variances With the \textit{F}-Test}
\label{app:comp_pair_var_w_f_test}

We start with a set of paired response variances for the corresponding MOS regions from a pair of experiments. Since, to the best of our knowledge, there is no statistical method to compare a set of paired variances on a global level (i.e., family-wise), we resort to testing each pair of variances individually, and apply a correction for the multiple tests conducted. To compare a single pair of variances, typically the $F$-test is used. This test considers the ratio of variances as its test statistic:
$$F_{stat} = \frac{VAR_1}{VAR_2},$$
with $VAR_1 \geq VAR_2$, and switched experiment order otherwise. However, MOS regions with a response variance of 0, where all participants chose the same category, e.g., at a MOS of 5.0, break the test statistic. To overcome the problem, we introduce a threshold making sure that only MOS regions with variances above 0.1 are considered significantly different from MOS regions with variance equal to 0. $F_{stat}$ follows an $F$ distribution with $n_1-1$ and $n_2-1$ degrees of freedom, where $n_1$ and $n_2$ correspond to the number of responses used to estimate response variance for this MOS region in experiment 1 and experiment 2, respectively. To find a $p$-value for a single variance pair of interest, we use the following formula:
\begin{equation}
    p=P(F(n_1-1,n_2-1)\geq F_{stat}).
\end{equation}
As $VAR_1 \geq VAR_2$ forced right-tailed test, the obtained $p$-value has to be multiplied by 2 to obtain the $p$-value of the two-tailed test.

Finally, as we are conducting multiple $F$-tests (i.e., one per each MOS region considered), we are facing the multiple comparisons problem, i.e., the more tests are conducted, the higher the probability of erroneous rejection by chance. Thus, the significance thresholds of the individual tests have to be more strict. This means that $p$-values need to be corrected, e.g., by Bonferroni correction or Holm-Sidak adjustment. Having applied the $p$-value correction, we compare the adjusted $p$-values to the global significance threshold of 0.05 (or any other assumed significance level). If the smallest adjusted $p$-value is less than 0.05, the experiments cannot be considered to provide similar variances for all MOS regions. This means there is at least one MOS region for which the difference in response variance is significant.

As a final remark, note that the $F$-test requires normally distributed data. This typically does not hold for response distributions in QoE experiments. Thus, Levene’s test would be better suited here. However, we failed to find a test statistic for MOS regions, which were only tested in one experiment. This is because the test statistic in Levene's test is based on mean absolute error (MAE) and not variance. To use it, we would need to fit the MOS--MAE curve instead of the MOS--SOS curve. Naturally, to overcome this limitation, we could limit the comparisons to MOS regions that are covered in both experiments. However, depending on the stimuli tested in the compared experiments, the overlap of MOS regions could be small. In turn, this would reduce the meaningfulness of the result this approach yields.


\section{Data Generator Configuration Influence on Responses Generated}
\label{app:data_gen_conf_influence}
\begin{figure}[!t]
    \includegraphics[width=0.85\columnwidth]{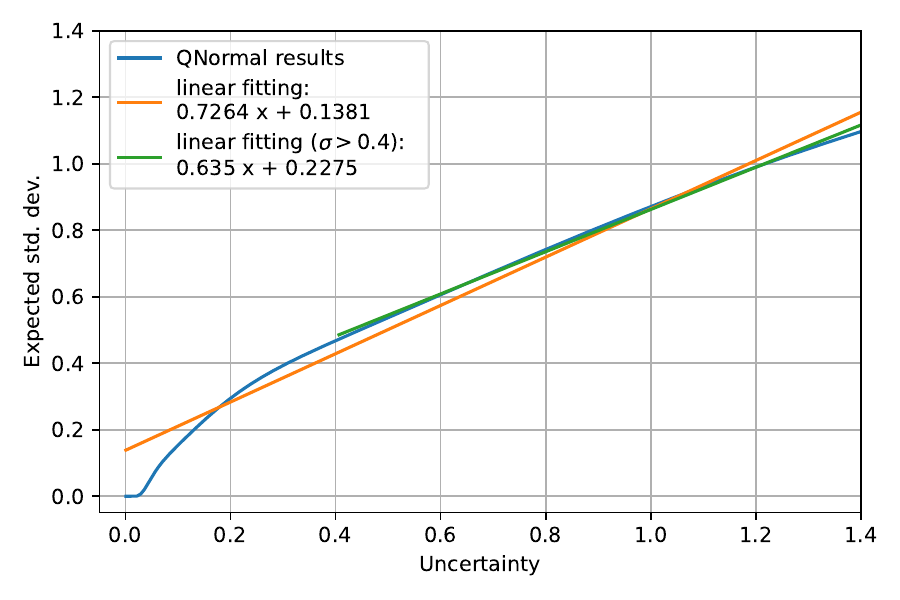}
    \caption{Expected standard deviation of responses generated depending on uncertainty $\sigma$ of users for the $\mathrm{QNorm}$ model (cf. (\ref{eq:qnorm})).}
    \label{fig:uncertainty2stdev}
\end{figure}

Figure~\ref{fig:uncertainty2stdev} illustrates one effect of discretisation and censoring (inherent to the $\mathrm{QNorm}$ model) that influences how what we provide to the data generator translates into the responses generated.
We consider a nonbiased user with some uncertainty $\sigma$. Then, we numerically derive, for any stimulus $x=1,2,\dots, k$, the resulting standard deviation $\widetilde{\sigma}_x = \STD{Q_x}$  of the user rating distribution.
Distribution shape is defined by parameters $\mu_x$ and $\sigma$. Hence, user responses for stimulus $x$ (denoted as a random variable $Q_x$) follow the $\mathrm{QNorm}(\mu_x, \sigma)$ distribution.
The formula for the mean standard deviation $\bar{\sigma}$ (over the $k$ stimuli) of the responses generated from this distribution is in~(\ref{eq:mean_sigma}). 

Fig.~\ref{fig:uncertainty2stdev} shows how $\bar{\sigma}$ is changing depending on the $\sigma$ value provided to the data generator. In this figure, we also show the linear fit between $\sigma$ and $\bar{\sigma}$. We do the fitting twice: once for the entire range of $\sigma$ values tested and once for $\sigma > 0.4$.
We can see that there is an almost linear dependency between $\sigma$ and $\bar{\sigma}$ for $\sigma>0.4$. Hence, for $\sigma > 0.4$, we assume that changes we make to data generator's $\sigma$ translate into linear changes in the expected standard deviation of the responses generated.
\begin{align}
& \bar{\sigma} = \frac{1}{k}\sum_{x=1}^k \widetilde{\sigma}_x = \frac{1}{k}\sum_{x=1}^k \E{Q_x^2} - \E{Q_x}^2 \label{eq:mean_sigma}
\shortintertext{with}
& \E{Q_x} = \sum_{j=1}^5 j \cdot P(Q_x=j) \;, \nonumber\\
& \E{Q_x^2} = \sum_{j=1}^5 j^2 \cdot P(Q_x=j) \;.\nonumber
\end{align}

\begin{figure}[!t]
    \centering
    \includegraphics[width=\columnwidth]{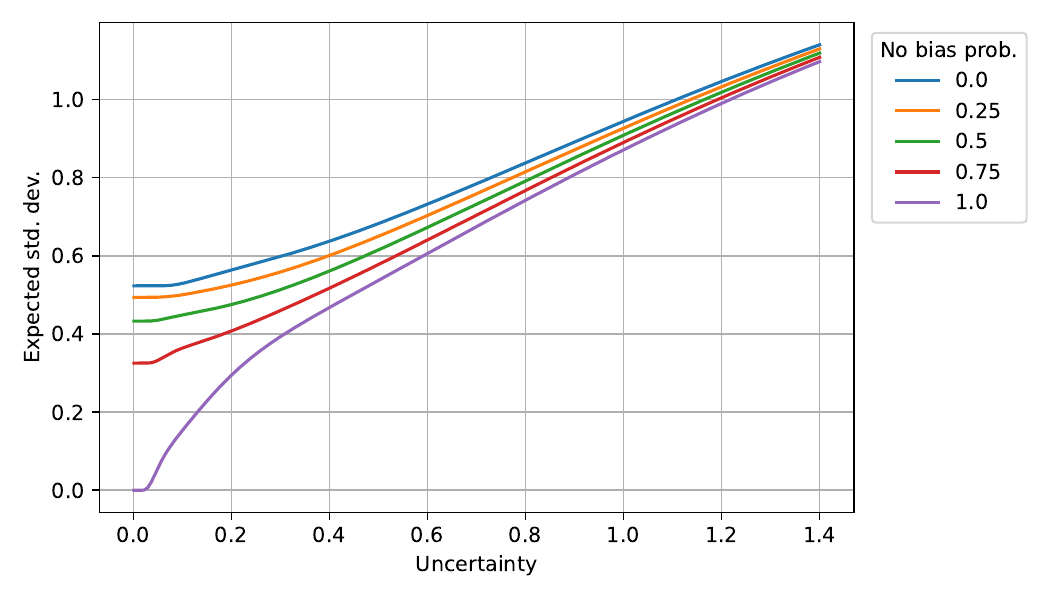}
    \caption{Expected standard deviation of responses generated for the mixed symmetric bias scenario depending on uncertainty $\sigma$ of users and the no bias probability $p$.}
    \label{fig:uncertaintyMixed}
\end{figure}

Figure~\ref{fig:uncertaintyMixed} additionally shows how the expected standard deviation of the responses generated ($\bar{\sigma}$) behaves in the mixed symmetric bias scenario with different values of the no bias probability $p$. We can see that the greater is the number of biased users (i.e., the \textit{lower} is the value of the $p$ parameter), the greater expected standard deviation of the responses generated we observe. This is expected as introducing a symmetric bias artificially increases the spread of the responses generated. Again, this confirms that our data generation strategy works properly.


\section{Confidence Intervals Computation Procedure}
\label{app:confidence_intervals}
In Figures~\ref{fig:rho_bar_vs_no_bias_prob}, \ref{fig:sosa_vs_no_bias_prob}, and \ref{fig:upsilon_bar_vs_no_bias_prob} we show 95\% confidence intervals for the measure $g$, $a$, and $\ell$, respectively. In this section, we make it clear how we compute these confidence intervals.

Let us focus on the $g$ measure's confidence interval. Let us assume that we have a simulated experiment with 21 stimuli. For each stimulus, we need to fit the GSD model. Having done that, we have a vector of 21 estimated $\rho$s (since this is the only GSD model parameter that we consider in this paper). Let us denote $\rho$ estimated for the $i$-th stimulus as $\hat{\rho_i}$. Now, having a vector of 21 $\hat{\rho_i}$s, we can compute two basic summary statistics that describe the mean (denoted as $\bar{\hat{\rho}}$) and the standard deviation (denoted as $\text{std}(\hat{\rho})$). Please note that $\bar{\hat{\rho}} = g$ (cf. Sec.~\ref{ssec:gsd_based_measure}).

Since we repeat the data generation step for the same experiment configuration (i.e., a fixed pair of no bias probability $p$ and unconfidence $\sigma$) multiple times (e.g., $R=200$ times), we get $R$ values of $\bar{\hat{\rho}}$ and $R$ values of $\text{std}(\hat{\rho})$. Let us denote the mean and standard deviation of estimated $\hat{\rho}$s for the $r$-th repetition of a simulated experiment as $\bar{\hat{\rho}}^r$ and $\text{std}(\hat{\rho})^r$, respectively.

Now, although we repeat the simulation $R$ times, we want to have one 95\% confidence interval.
This confidence interval should give an idea of how the $g$ measure varies when we repeat the same simulated experiment multiple times. 
For this, we certainly need the average $g$ (averaged over $R$ simulation runs). Let us denote this global average as $\bar{g}$. Next, we also need the average standard deviation $\overline{\text{std}(\hat{\rho})}$. To find it, we use the following formula:
\begin{equation*}
   \overline{\text{std}(\hat{\rho})} = \frac{1}{R} \sum_{r=1}^R \text{std}(\hat{\rho})^r.
\end{equation*}

Finally, we can compute the 95\% confidence interval for $g$ (in the context of repeating each simulated experiment $R$ times):
\begin{equation} \label{eq:g_conf_int}
    \bar{g} \pm t_{\alpha/2, K-1} \cdot \frac{\overline{\text{std}(\hat{\rho})}}{\sqrt{K-1}}, 
\end{equation}
where $K$ stands for the number of stimuli in a single simulated experiment ($K=21$ throughout this paper); $\alpha$ is the significance level (5\% in our case) and $t_{\alpha/2, \nu}$ is the $100\cdot\alpha/2$ percentage point of the $t$-distribution with $\nu$ degrees of freedom.

Please note that the width of the confidence interval given in (\ref{eq:g_conf_int}) does \textit{not} depend on the number of repetitions $R$ of each simulated experiment. We consider this to be a desirable feature of the method as it does not give a false impression that $g$ can be estimated with an arbitrary precision. In other words, this confidence interval gives a good intuition regarding how precise the estimation of $g$ is for a real experiment with $K$ stimuli, while still providing the added benefit of stable average $g$ estimation due to the high number of simulation runs.

If it comes to the $\ell$ measure, the procedure is very similar to the one given for the $g$ measure. The only difference is that the $\ell$ measure uses per subject information and so $K$ now means the number of subjects taking part in the experiment considered. Naturally, in the case of the $\ell$ measure, we are also interested in estimating the $\upsilon$ parameter of the Li2020 model (and not $\rho$).

For the $a$ measure, the treatment is again similar to the one presented in (\ref{eq:g_conf_int}). However, this time we multiply $t_{\alpha/2, K-1}$ by the average standard error of the estimator of the measure $a$. To be more specific, when estimating the value of the $a$ measure for each simulated experiment, we get both the $a$ measure and the standard error of its estimator (let us denote it by $\mathrm{SE}(a)$). Since we have $R$ repetitions of each simulated experiment, we need to first calculate the average standard error $\overline{\mathrm{SE}(a)}$ before we can use it to find the confidence interval. In general, we use the following formula to find the confidence interval for the measure $a$:
\begin{align*}
    \bar{a} \pm t_{\alpha/2, K-1} \cdot \overline{\mathrm{SE}(a)},
\end{align*}
where $\bar{a}$ is the average measure $a$ (averaged over $R$ runs of a simulated experiment of interest).


\section{Justification for Selecting $\upsilon$ as a Measure of Experiment Precision}
\label{app:justification}

The Li2020 model~\cite{Li2020simple} has three parameters: i) subject bias $\Delta$, ii) true quality $\psi$, and iii) subject inconsistency $\upsilon$. In Sec.~\ref{ssec:li2020_based_measure} we decided to use the subject inconsistency parameter $\upsilon$ as a basis for our precision measure $\ell$. The following text justifies our approach.

Although one could argue that subject bias is somehow related to experiment precision, in Sec.~\ref{ssec:notion_of_experiment_precision} we explicitly say that we assume that subject bias does \emph{not} influence experiment precision. Naturally, one can disagree with this assumption. However, with this assumption in place, it would be pointless to use the subject bias parameter of the Li2020 model as a measure of experiment precision.

When it comes to the true quality parameter $\psi$, it is very similar to the MOS. In other words, it expresses the mean opinion of human observers. Hence, this parameter does \emph{not} express the variance of subjective responses. Without any link to response variance, this parameter is a very poor candidate for the experiment precision assessment measure.

The subject inconsistency parameter $\upsilon$ of the Li2020 model is the only parameter of the model designed to express the response variance (excluding the variance resulting from the subject's bias). This fact alone makes it a good candidate for a measure of experiment precision.

Finally, the measure $\ell$ is simply an average subject inconsistency. We take the average since we are interested in the experiment-wise precision and not the precision of each subject individually.

\begin{figure}[!t]
    \centering
    \resizebox{0.31\textwidth}{!}{
\begin{tikzpicture}

\definecolor{color0}{rgb}{0.886274509803922,0.290196078431373,0.2}
\definecolor{color1}{rgb}{0.203921568627451,0.541176470588235,0.741176470588235}
\definecolor{color2}{rgb}{0.596078431372549,0.556862745098039,0.835294117647059}
\definecolor{color3}{rgb}{0.984313725490196,0.756862745098039,0.368627450980392}
\definecolor{color4}{rgb}{0.556862745098039,0.729411764705882,0.258823529411765}
\definecolor{color5}{rgb}{1,0.709803921568627,0.72156862745098}

\begin{axis}[
axis background/.style={fill=white},
axis line style={white!89.8039215686275!black},
tick align=outside,
tick pos=left,
x grid style={white!89.8039215686275!black},
xlabel={MOS},
xmajorgrids,
xmin=0.8, xmax=5.2,
xtick style={color=white!33.3333333333333!black},
y grid style={white!89.8039215686275!black},
ylabel={Variance},
ymajorgrids,
ymin=-0.200000000000015, ymax=4.2,
ytick style={color=white!33.3333333333333!black},
ytick={-0.5,0,0.5,1,1.5,2,2.5,3,3.5,4,4.5},
yticklabels={\ensuremath{-}0.5,0.0,0.5,1.0,1.5,2.0,2.5,3.0,3.5,4.0,4.5}
]
\path [draw=blue, fill=blue, opacity=0.15, very thin]
(axis cs:1,0)
--(axis cs:1,0)
--(axis cs:1.05,0.0475)
--(axis cs:1.1,0.0900000000000001)
--(axis cs:1.15,0.1275)
--(axis cs:1.2,0.16)
--(axis cs:1.25,0.1875)
--(axis cs:1.3,0.21)
--(axis cs:1.35,0.2275)
--(axis cs:1.4,0.24)
--(axis cs:1.45,0.2475)
--(axis cs:1.5,0.25)
--(axis cs:1.55,0.2475)
--(axis cs:1.6,0.24)
--(axis cs:1.65,0.2275)
--(axis cs:1.7,0.21)
--(axis cs:1.75,0.1875)
--(axis cs:1.8,0.16)
--(axis cs:1.85,0.127499999999999)
--(axis cs:1.9,0.0899999999999994)
--(axis cs:1.95,0.0474999999999992)
--(axis cs:2,8.88178419700124e-16)
--(axis cs:2.05,0.0475000000000006)
--(axis cs:2.1,0.0900000000000008)
--(axis cs:2.15,0.127500000000001)
--(axis cs:2.2,0.160000000000001)
--(axis cs:2.25,0.1875)
--(axis cs:2.3,0.21)
--(axis cs:2.35,0.2275)
--(axis cs:2.4,0.24)
--(axis cs:2.45,0.2475)
--(axis cs:2.5,0.25)
--(axis cs:2.55,0.2475)
--(axis cs:2.6,0.24)
--(axis cs:2.65,0.2275)
--(axis cs:2.7,0.209999999999999)
--(axis cs:2.75,0.187499999999999)
--(axis cs:2.8,0.159999999999999)
--(axis cs:2.85,0.127499999999999)
--(axis cs:2.9,0.0899999999999987)
--(axis cs:2.95,0.0474999999999982)
--(axis cs:3,1.77635683940025e-15)
--(axis cs:3.05,0.0475000000000014)
--(axis cs:3.1,0.0900000000000015)
--(axis cs:3.15,0.127500000000002)
--(axis cs:3.2,0.160000000000001)
--(axis cs:3.25,0.187500000000001)
--(axis cs:3.3,0.210000000000001)
--(axis cs:3.35,0.227500000000001)
--(axis cs:3.4,0.24)
--(axis cs:3.45,0.2475)
--(axis cs:3.5,0.25)
--(axis cs:3.55,0.2475)
--(axis cs:3.6,0.24)
--(axis cs:3.65,0.227499999999999)
--(axis cs:3.7,0.209999999999999)
--(axis cs:3.75,0.187499999999999)
--(axis cs:3.8,0.159999999999999)
--(axis cs:3.85,0.127499999999998)
--(axis cs:3.9,0.0899999999999979)
--(axis cs:3.95,0.0474999999999974)
--(axis cs:4,2.66453525910037e-15)
--(axis cs:4.05,0.0475000000000022)
--(axis cs:4.1,0.0900000000000026)
--(axis cs:4.15,0.127500000000002)
--(axis cs:4.2,0.160000000000002)
--(axis cs:4.25,0.187500000000001)
--(axis cs:4.3,0.210000000000001)
--(axis cs:4.35,0.227500000000001)
--(axis cs:4.4,0.240000000000001)
--(axis cs:4.45,0.2475)
--(axis cs:4.5,0.25)
--(axis cs:4.55,0.2475)
--(axis cs:4.6,0.239999999999999)
--(axis cs:4.65,0.227499999999999)
--(axis cs:4.7,0.209999999999999)
--(axis cs:4.75,0.187499999999998)
--(axis cs:4.8,0.159999999999998)
--(axis cs:4.85,0.127499999999998)
--(axis cs:4.9,0.0899999999999969)
--(axis cs:4.95,0.0474999999999966)
--(axis cs:5,3.55271367880049e-15)
--(axis cs:5,-1.4210854715202e-14)
--(axis cs:5,-1.4210854715202e-14)
--(axis cs:4.95,0.197499999999985)
--(axis cs:4.9,0.389999999999985)
--(axis cs:4.85,0.577499999999988)
--(axis cs:4.8,0.759999999999988)
--(axis cs:4.75,0.937499999999988)
--(axis cs:4.7,1.10999999999999)
--(axis cs:4.65,1.27749999999999)
--(axis cs:4.6,1.43999999999999)
--(axis cs:4.55,1.59749999999999)
--(axis cs:4.5,1.74999999999999)
--(axis cs:4.45,1.89749999999999)
--(axis cs:4.4,2.03999999999999)
--(axis cs:4.35,2.17749999999999)
--(axis cs:4.3,2.30999999999999)
--(axis cs:4.25,2.43749999999999)
--(axis cs:4.2,2.55999999999999)
--(axis cs:4.15,2.67749999999999)
--(axis cs:4.1,2.78999999999999)
--(axis cs:4.05,2.89749999999999)
--(axis cs:4,2.99999999999999)
--(axis cs:3.95,3.09749999999999)
--(axis cs:3.9,3.19)
--(axis cs:3.85,3.2775)
--(axis cs:3.8,3.36)
--(axis cs:3.75,3.4375)
--(axis cs:3.7,3.51)
--(axis cs:3.65,3.5775)
--(axis cs:3.6,3.64)
--(axis cs:3.55,3.6975)
--(axis cs:3.5,3.75)
--(axis cs:3.45,3.7975)
--(axis cs:3.4,3.84)
--(axis cs:3.35,3.8775)
--(axis cs:3.3,3.91)
--(axis cs:3.25,3.9375)
--(axis cs:3.2,3.96)
--(axis cs:3.15,3.9775)
--(axis cs:3.1,3.99)
--(axis cs:3.05,3.9975)
--(axis cs:3,4)
--(axis cs:2.95,3.9975)
--(axis cs:2.9,3.99)
--(axis cs:2.85,3.9775)
--(axis cs:2.8,3.96)
--(axis cs:2.75,3.9375)
--(axis cs:2.7,3.91)
--(axis cs:2.65,3.8775)
--(axis cs:2.6,3.84)
--(axis cs:2.55,3.7975)
--(axis cs:2.5,3.75)
--(axis cs:2.45,3.6975)
--(axis cs:2.4,3.64)
--(axis cs:2.35,3.5775)
--(axis cs:2.3,3.51)
--(axis cs:2.25,3.4375)
--(axis cs:2.2,3.36)
--(axis cs:2.15,3.2775)
--(axis cs:2.1,3.19)
--(axis cs:2.05,3.0975)
--(axis cs:2,3)
--(axis cs:1.95,2.8975)
--(axis cs:1.9,2.79)
--(axis cs:1.85,2.6775)
--(axis cs:1.8,2.56)
--(axis cs:1.75,2.4375)
--(axis cs:1.7,2.31)
--(axis cs:1.65,2.1775)
--(axis cs:1.6,2.04)
--(axis cs:1.55,1.8975)
--(axis cs:1.5,1.75)
--(axis cs:1.45,1.5975)
--(axis cs:1.4,1.44)
--(axis cs:1.35,1.2775)
--(axis cs:1.3,1.11)
--(axis cs:1.25,0.937500000000001)
--(axis cs:1.2,0.760000000000001)
--(axis cs:1.15,0.5775)
--(axis cs:1.1,0.39)
--(axis cs:1.05,0.1975)
--(axis cs:1,0)
--cycle;

\end{axis}

\end{tikzpicture}}
    \caption{A set of all possible MOS--variance pairs. True for subjective responses expressed on a 5-level discrete response scale (e.g., the 5-level ACR scale).}
    \label{fig:ghost_like_shape}
\end{figure}
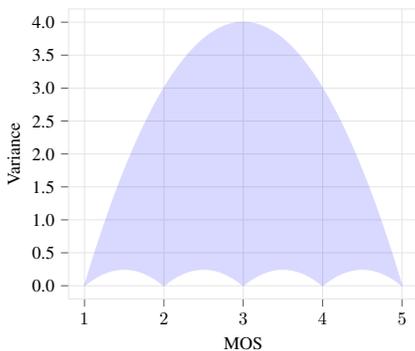


%








\end{document}